\PassOptionsToPackage{pdfpagelabels=false}{hyperref} 
\documentclass[fleqn,usenatbib]{mnras}

\usepackage{newtxtext,newtxmath}

\usepackage[T1]{fontenc}
\usepackage{ae,aecompl}

\usepackage[usenames, dvipsnames]{color}
\usepackage{graphicx}	

\newcommand{\orcid}[1]{\href{https://orcid.org/#1}{\includegraphics[width=10pt]{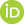}}}

\title[{\em Gaia} Preliminary SPSS flux tables]{The {\em Gaia} spectrophotometric standard stars survey -- V. Preliminary flux tables for the calibration of {\em Gaia} DR2 and (E)DR3}
\author[E. Pancino et al.]{E.~Pancino$^{1,2}$\thanks{E-mail: elena.pancino@inaf.it}\orcid{0000-0003-0788-5879},
N.~Sanna$^{1}$\orcid{0000-0001-9275-9492},
G.~Altavilla$^{3,2}$\orcid{0000-0002-9934-1352},
S.~Marinoni$^{3,2}$\orcid{0000-0001-7990-6849},
M.~Rainer$^{1}$\orcid{0000-0002-8786-2572},
G.~Cocozza$^{4}$,
\newauthor
S.~Ragaini$^{4}$,
S.~Galleti$^{4}$,
M.~Bellazzini$^{4}$\orcid{0000-0001-8200-810X},
A.~Bragaglia$^{4}$\orcid{0000-0002-0338-7883},
G.~Tessicini$^{4}$,
H.Voss$^{5}$,
J.~M.~Carrasco$^{5}$\orcid{0000-0002-3029-5853},
\newauthor
C.~Jordi$^{5}$\orcid{0000-0001-5495-9602},
D.~L.~Harrison$^{6,7}$\orcid{0000-0001-8687-6588},
F.~De~Angeli$^{7}$\orcid{0000-0003-1879-0488},
D.~W.~Evans$^{7}$\orcid{0000-0002-6685-5998},
G.~Fanari$^{2}$
\vspace{0.3cm}\\
$^{1}$ INAF -- Osservatorio Astrofisico di Arcetri, Largo E. Fermi 5, I-50125, Florence, Italy\\
$^{2}$ Space Science Data Center -- ASI, Via del Politecnico SNC, I-00133 Roma, Italy\\
$^{3}$ INAF -- Osservatorio Astronomico di Roma, Via Frascati 33, I-00078, Monte Porzio Catone (Roma), Italy \\
$^{4}$ INAF -- Osservatorio di Astrofisica e Scienza dello Spazio, via Gobetti 93/3, I-40129 Bologna, Italy\\
$^{5}$ Departament de F\'{i}sica Qu\`{a}ntica i Astrof\'{i}sica, 
Institut de Ci\`encies del Cosmos (ICCUB),  Universitat de Barcelona (IEEC-UB), Mart\'\i\  i Franqu\`es, 1,\\
\,\,                E-08028 Barcelona, Spain\\ 
$^{6}$ Kavli Institute for Cosmology, Institute of Astronomy, Madingley Road, Cambridge, CB3 0HA, UK\\
$^{7}$ Institute of Astronomy, University of Cambridge, Madingley Road, Cambridge CB3 0HA, UK\\
}
\date{Accepted XXX. Received YYY; in original form ZZZ}

\pubyear{2021}

\begin{document}
\label{firstpage}
\pagerange{\pageref{firstpage}--\pageref{lastpage}}
\maketitle

\begin{abstract}
We present the flux tables of the spectro-photometric standard stars (SPSS) used to calibrate in flux the {\em Gaia} DR2 and (E)DR3 data releases. The latest SPSS grid version contains 112 stars, whose flux tables agree to better than 1\% with the CALSPEC spectra of 11 flux standards for the calibration of the Hubble Space Telescope. The synthetic magnitudes computed on the SPSS spectra also agree to better than 1\% with the Landolt magnitudes of 37 stars in common. The typical spreads in both comparisons are of the order of 1\%. These uncertainties already meet the initial requirements for the {\em Gaia} SPSS project, but further improvements are expected in the next SPSS versions, that will be used to calibrate future {\em Gaia} releases. We complement the SPSS flux tables with literature spectra of 60 additional stars that did not pass all the criteria to be SPSS, the Passband Validation Library (PVL). The PVL contains stars of extreme spectral types, such as bright O and B stars and late M stars and brown dwarfs, and was useful to investigate systematic effects in the previous {\em Gaia} DR2 release and to minimize them in the EDR3 one. The PVL literature spectra are recalibrated as accurately as possible onto the SPSS reference scale, so that the two sets together can be used in a variety of validation and comparison studies. 
\end{abstract}



%

\section{Introduction}
\label{sec:intro}

The ESA astrometric mission {\em Gaia}\footnote{\url{https://www.cosmos.esa.int/web/gaia}} \citep{gaia} has revolutionized our view of the Milky Way and its data are bringing progress in almost every area of astrophysical research \citep{perryman01,gilmore18,pancino20}. Three data releases were published so far\footnote{\url{https://www.cosmos.esa.int/web/gaia/release}}, with a constant increase in the data quality and in the quantity of data products \citep{gdr1,gdr2,egdr3}. Several thousand publications in different areas of astrophysical research cite or use {\em Gaia} data, from solar system objects and fundamental physics to the study of stars, stellar clusters, the Milky Way, and the Local Group. {\em Gaia} also tackles time-variability phenomena and provides large samples of external galaxies and distant quasars. The EDR3 (Early Data Release 3) {\em Gaia} catalogue contains astrometry, i.e., positions, parallaxes, and proper motions for almost two billions of stars and non-stellar sources, down to a magnitude of G$\simeq$21~mag \citep[V$\simeq$20.5, see][]{riello20}. The quality of the {\em Gaia} astrometry is unprecedented: errors on the EDR3 measurements are of the order of 10--100~$\mu$as \citep[20--130~$\mu$uas on parallaxes and 20-140~$\mu$as/yr on proper motions, see][]{lindegren20}. The data are complemented by three-band photometry \citep{riello20} and the upcoming DR3 (Data Release 3) will present low-resolution spectra covering the optical range (330--1050~nm). Line-of-sight velocities are also presented for more than seven million stars down to G$\lesssim$14~mag \citep{katz19}, thanks to higher resolution (R=$\lambda/\delta\lambda\simeq$11\,500) spectra in the Calcium triplet region, and the {\em Gaia} data are complemented by stellar parameter estimates \citep{andrae18}. The multi-epoch observations of {\em Gaia} allow for time-domain studies as well \citep{eyer19}.

The complex {\em Gaia} data analysis system, developed within the Data Processing and Analysis Consortium (DPAC), provides essentially self-calibrated data products. However, for some deliverables it is necessary to rely on external catalogues to bring the {\em Gaia} measurements onto an external physical system of standardized quantities. This is for example the case of the radial (or line-of-sight) velocities, where a specific set of thousands of carefully selected standard stars, constancy monitored to be stable within $\simeq$300~m~s$^{-1}$, was assembled for the specific purpose of calibrating {\em Gaia} high-resolution spectra \citep{soubiran18}. The external calibration is even more necessary in the case of fluxes and magnitudes, because {\em Gaia} counts electrons like any typical modern astronomical detector, and thus an external calibration is necessary to transform the measurements into a physical scale. Unfortunately, only a few primary flux standards exist in the literature, such as Vega or Sirius (see also Section~\ref{sec:concl} for more details), and they are too bright for {\em Gaia}, which is not designed for bright stars and is incomplete above 3--6~mag. The most suitable and reliable sets of secondary standards to calibrate {\em Gaia} fluxes are the CALSPEC grid of spectrophotometric standard stars\footnote{\url{https://www.stsci.edu/hst/instrumentation/reference-data-for-calibration-and-tools/astronomical-catalogs/calspec}} used to calibrate the Hubble Space Telescope \citep[HST,][]{bohlin14} and the Landolt standards for the Johnson-Kron-Cousins broadband photometric system \citep[][see also Section~\ref{sec:landolt}]{landolt92}. We thus decided to tie the {\em Gaia} flux calibration to the three pure hydrogen white dwarfs (WD) used as reference in CALSPEC, that we adopted as the pillars of our calibration \citep{pancino12}. However, {\em Gaia} is a sophisticated space observatory, capable of delivering photometry with a precision of a few mmag. To reach the best possible accuracy, the external calibration model of {\em Gaia} requires a large set of Spectro-Photometric Standard Stars (SPSS) subject to well-defined requirements \citep[][see also Section~\ref{sec:sele}]{pancino12}. Unfortunately, the available standard grids did not contain the required number of stars (about 200) meeting our requirements. We thus started a new set of ground-based observing campaigns to build a specific {\em Gaia} SPSS grid, calibrated on the flux system defined by CALSPEC.

This is the fifth paper of the {\em Gaia} SPSS series, preceded by: a presentation of the SPSS survey and some preliminary results \citep{pancino12}; the description of the data reduction methods and the characterization of the instruments used \citep{altavilla15}; the results of the constancy monitoring of SPSS candidates \citep{marinoni16}; and the presentation of the calibrated magnitudes of candidate SPSS \citep{altavilla20}. We present here two preliminary versions of the SPSS flux tables. The SPSS V1 was used to calibrate the photometry in the second {\em Gaia} release \citep[DR2,][]{gdr2,evans18}; the SPSS V2 was used to calibrate the photometry in EDR3 \citep{egdr3,riello20} and will be used to calibrate the low-resolution spectra in DR3, expected in the first half of 2022. We complement the paper by presenting also the Passband Validation Library (PVL), that was built with the purpose of investigating systematic effects in the {\em Gaia} photometry and passband reconstruction, and more in general as a validation sample. The PVL contains stars of more extreme spectral types that did not fulfil all the criteria to be selected as SPSS (Section~\ref{sec:sele}) and is based on literature data. 

The paper is organized as follows: in Section~\ref{sec:obs} we discuss the SPSS selection criteria and the observing campaigns; in
Section~\ref{sec:data} we present the procedures employed for the spectra reductions; in Section~\ref{sec:calib} we describe the flux calibration procedure and the flux tables preparation; in Section~\ref{sec:res} we describe the SPSS results and validate them against other literature samples; in Section~\ref{sec:pvl} we present the PVL; and in Section~\ref{sec:concl} we summarize our results and discuss them in the context of the existing data-sets and future spectro-photometric projects.

\begin{table*}
\caption{List of SPSS included in the V1 and V2 sets along with basic information: the SPSS ID and name; the {\em Gaia} EDR3 ID and coordinates (rounded in the pdf table version); the B and V magnitudes from \citet{altavilla20}, except for LTT\,377 (SPSS\,350), where they are from \citet{koen10}; the spectral type collected from the literature by \citet{pancino12}; the SPSS set in which the star was included, either V1 for {\em Gaia} DR2 or V2 for {\em Gaia} (E)DR3; the quality flag (0 = Ok, 1 = warning on minor spectral defects, 2 = warning on spectral defects) and the notes. Only the first five lines are shown here; the table is available in its entirety online (see Data Availability Section). }\label{tab:spsslist}
\begin{tabular}{|l|l|l|r|r|r|r|r|r|r|r|l|}
\hline
ID &  Name   & {\em Gaia} EDR3 ID & RA & Dec & B & V & SpType & V1 & V2 & Flag & Notes \\
& & & (deg) & (deg) & (mag) & (mag) &  & ---\\
\hline                                   001     & G191-B2B & 266077145295627520  & 76.37767 &  52.83067 & 11.44 & 11.79 & DA0 & yes & yes & 0 & ---\\
002     & GD\,71   & 3348071631670500736 &   88.11544 & 15.88624 & 12.79 & 13.03 & DA1 & yes & yes & 0 & --- \\
003     & GD\,153  & 3944400490365194368 & 194.25949 &  22.03039 & 13.11 & 13.40 & DA1 & yes & yes & 0 & --- \\
005     & EG\,21   & 4646535078125821568 &   47.62973 & --68.60140 & 11.40 & 11.40 & DA3 & yes & yes & 0 & --- \\
006     & GD\,50   & 3251244858154433536 &   57.20948 & --0.97636 & 13.79 & 14.07 & DA2 & yes & yes & 0 & --- \\
\hline
\end{tabular}
\end{table*}

\begin{table*}
\caption{Summary of the facilities used for the SPSS spectroscopy observations. For each facility, we report the number of used spectra (for V1 and V2) on the total spectra acquired with the facility, the number of used observing nights on the total nights awarded, and the name and the wavelength coverage (in nm) of the used grisms. Note that the wheather conditions in Calar Alto and Loiano were rarely fully photometric (see Section~\ref{sec:calib}), thus the spectra obtained with these facilities will mostly be used in a future SPSS release to increase the S/N ratio and to smooth out pixel-to-pixel defects for SPSS observed with NTT and TNG.} \label{tab:tels}
\begin{tabular}{|l|l|r|r|r|r|r|r|r|}
\hline
Instrument          & Site              & Spectra (used/all) & Nights (used/awarded) &  blue grism       & red grism & third grism \\
\hline                                                          
EFOSC2@NTT\,3.6m    & La Silla, Chile   &         896 / 2037 &               17 / 42 &    \#11 (338--752) & \#16 (602--1032) & \#5 (520--935) \\
DOLoRes@TNG\,3.6m   & La Palma, Spain   &         650 / 2487 &                8 / 59 &    LR-B (300--843) & LR-R (447--1007) & &\\
CAFOS@2.2m          & Calar Alto, Spain &           15 / 995 &                1 / 90 &    B200 (320--900) & R200 (620--1100) & &\\
BFOSC@Cassini\,1.5m & Loiano, Italy     &            0 / 947 &              
0 / 32 &     \#3 (330--642) &   \#5 (480--980) & &\\
\hline
\end{tabular}
\end{table*}

\section{SPSS selection and observations} 
\label{sec:obs}

\begin{figure} 
\includegraphics[clip,width=\columnwidth]{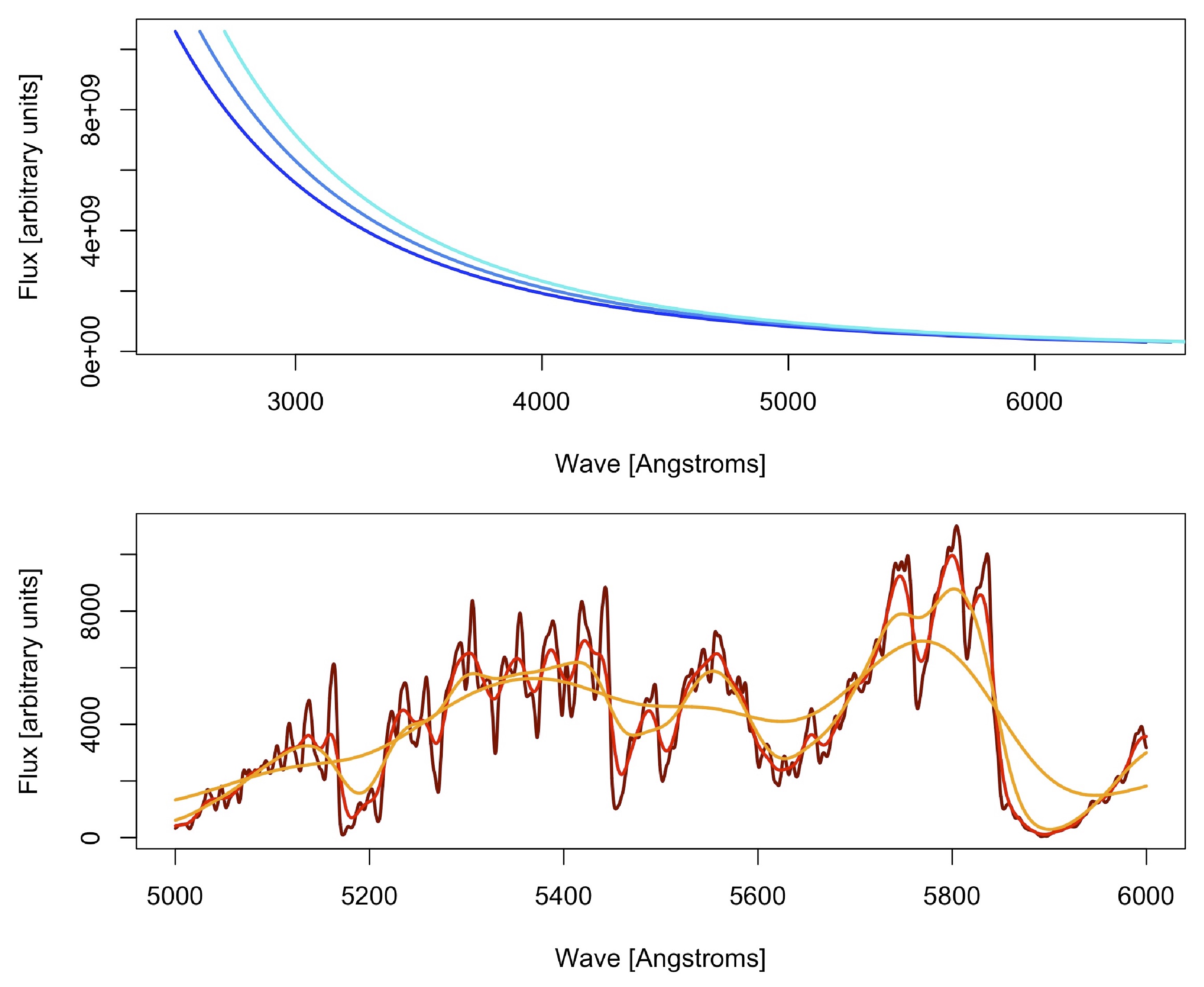} 
\vspace{-0.3cm}
\caption{Cartoon illustrating the risks of using certain spectral type stars as flux calibrators. {\em Top panel:} when using a featureless WD (or a very hot star), uncertainties in the wavelength calibration can imply large flux variations in the steepest blue spectral regions. Three black-body spectra (blue shades) are shifted by 10~nm, to make the effect clearly appreciable by eye (our uncertainties are of $\simeq$0.2~nm). {\em Bottom panel:} when using cool stars with deep molecular bands, variations in the effective spectral resolution, that can occur when using wide slits with variable seeing conditions, can imply large local flux variations. The mock spectra have resolution varying in a wide range from 0.5 (darkest) to 10~nm (lightest), to make the effect clearly appreciable visually.  Section~\ref{sec:sele}.}
\label{fig:danger}   
\end{figure}

\subsection{Candidate SPSS selection}
\label{sec:sele}

The selection criteria for the candidate SPSS stars were detailed in \citet{pancino12}. They were set to meet the {\em Gaia} calibration needs and concerned the resolution (R=$\lambda/\delta\lambda\simeq$1000, a few times the {\em Gaia} BP/RP resolution), S/N ratio ($>$100), magnitude (9$\lesssim$V$\lesssim$15~mag) distribution, color distribution (as wide as possible, see below), and sky distribution of the targets for the ground-based observations, having in mind that the stars should be observable with 2--4~m telescopes from both hemispheres. The {\em Gaia} requirements imposed the coverage of the entire wavelength range and a set of at least a hundred SPSS of different spectral types, to match the target of $\simeq$1\% precision and $\simeq$1--3\% accuracy in the resulting flux tables. This led to an initial selection of about 300 candidate SPSS. An additional requirement of short-term flux variations smaller than $\pm$0.5\% (an amplitude of $\pm$0.005~mag) was imposed and all SPSS candidate close enough to some instability strip were monitored for constancy, as described by \citet{marinoni16}. Not even the CALSPEC grid of flux standard met simultaneously all the above criteria and thus we started a dedicated spectro-photometric observing campaign.

The most difficult criterium to meet was the color and spectral type coverage. To allow for the best {\em Gaia} flux calibration, the SPSS needed to be as different as possible from each other, with different overall spectral energy distributions. This was not easy also when considering our constancy criterium, because most of the objects with emission lines tend to be variable (including hot stars and quasars). We also realized that featureless white dwarfs (WD), that are generally considered to be the best calibrators for obvious reasons \citep[see, e.g.,][]{moehler14}, are subject to large flux calibration uncertainties. In fact, even small wavelength calibration errors might induce large flux errors in the steep blue part of the spectrum (Figure~\ref{fig:danger}, top panel). These errors cannot be repaired using stellar features, which are absent, or sky emission and absorption lines, because we observed at low resolution and with wide slits. Similarly, cool stars with many molecular absorption bands of varying depth can have locally very uncertain flux levels (Figure~\ref{fig:danger}, bottom panel). The local flux in fact varies significantly with the unavoidable small resolution variations in wide slit observations, where the effective resolution is governed by the atmospheric seeing. 

Therefore, the initial list of about 200 good SPSS candidates contained a relatively small number of O and B stars, and no spectral types later than about M2 \citep{pancino12}. The list of SPSS used for the calibration of {\em Gaia} DR2 \citep{gdr2} and {\em Gaia} (E)DR3 \citep{egdr3} is a subset of that initial list with a handful of additions, and can be found in Table~\ref{tab:spsslist} along with relevant information.

\subsection{Observations}
\label{sec:subobs}

Observations started in the second half of 2006 and were completed in July 2015. Overall, we were awarded more than 5\,000 observing hours at seven different observing facilities, of which a large fraction was carried out in visitor mode. We carried out three main campaigns: (i) a constancy monitoring campaign, described in details in \citet{marinoni16}; (ii) an absolute photometry campaign, described in details in \citet{altavilla20}; and (iii) a spectro-photometric campaign, for which we present the first results here. We also started a fourth campaign, devoted to the constancy monitoring on longer timescales (about three years) \citep[see][and Section~\ref{sec:sea}]{pancino12}. However, we did not have the resources to carry it out to completion and we considered that {\em Gaia} itself will be able to help prune out any remaining long-term variable SPSS (Section~\ref{sec:sea}). In this paper, we present two versions of the SPSS flux tables library: the first (SPSS V1) was used to calibrate {\em Gaia} DR2 and contains 94 SPSS observed in the best nights up to 2013; the second (SPSS V2) was used to calibrate {\em Gaia} (E)DR3 and contained more SPSS (112)\footnote{The actual SPSS list used to calibrate {\em Gaia} EDR3 contained 113 stars, but G\,184-20 (SPSS\,140) was later rejected because of identification problems in our ground-based observations.} and more spectra for each SPSS, including spectra observed after we completed the V1 flux tables in 2013.   

The spectro-photometric observations were carried out using four different facilities, listed in Table~\ref{tab:tels}, along with some relevant information. All spectra were taken with wide slits, to minimize (differential) flux losses, but narrow slit spectra were also obtained to allow for a more accurate wavelength calibration (Section~\ref{sec:prered}). Three wide exposures and one narrow exposure were taken for each SPSS in each of two set-ups, a blue one and a red one, covering the entire {\em Gaia} wavelength range (about 330-1050~nm, see Table~\ref{tab:tels}). Each star was observed more than once, whenever possible, in different nights and with different facilities. For the flux calibrations we observed several times at different airmasses, a set of {\em Pillars} and {\em Primary} SPSS\footnote{As described in \citet{pancino12}, the Primary SPSS are well known stars in the literature, mostly from the CALSPEC database, that can be used as calibrators when the Pillars are not visible.} as the night flux calibrators \citep[see][and Section~\ref{sec:calib} for more details]{pancino12}. Daytime and twilight calibrations were taken daily (see next section for more details) and we performed several dedicated experiments to characterize the stability and quality of the calibration data for each facility \citep[see][and following sections for more details]{altavilla15}.

\section{SPSS data reductions} 
\label{sec:data}

\subsection{Basic spectra reductions}
\label{sec:prered}

Every night we obtained at least 10 bias frames and 5 spectroscopic flat field frames for each setup used during the night. Some dark frames were taken as part of our instrument familiarization plan \citep[IFP,][]{altavilla15}, but we found that a dark correction was not necessary. In the few nights in which good calibrations could not be obtained, we used the results of our IFP calibration frames stability study (with a threshold of $\pm$1\%) to decide whether it was safe to use the master frames obtained in other nights or it was better to reject the data for the entire night. We used \textsc{iraf}\footnote{\url{http://iraf.noao.edu}} \citep{irafold,iraf} to create and apply master frames for each night, including the bias, flat field, and illumination corrections. Whenever possible, we used the overscan strips to correct the bias level before applying the bias and flat corrections to the 2D spectroscopic frames. We also created bad pixel masks, roughly twice per year, and used them to exclude hot or dead pixels from the following reduction steps. An illumination correction, obtained with sky flats, was applied besides the usual master flat-field correction, obtained from lamp flats. More details on the pre-reduction and the quality control procedures can be found in \citet{altavilla15}. 

\begin{figure} 
\includegraphics[clip,width=\columnwidth]{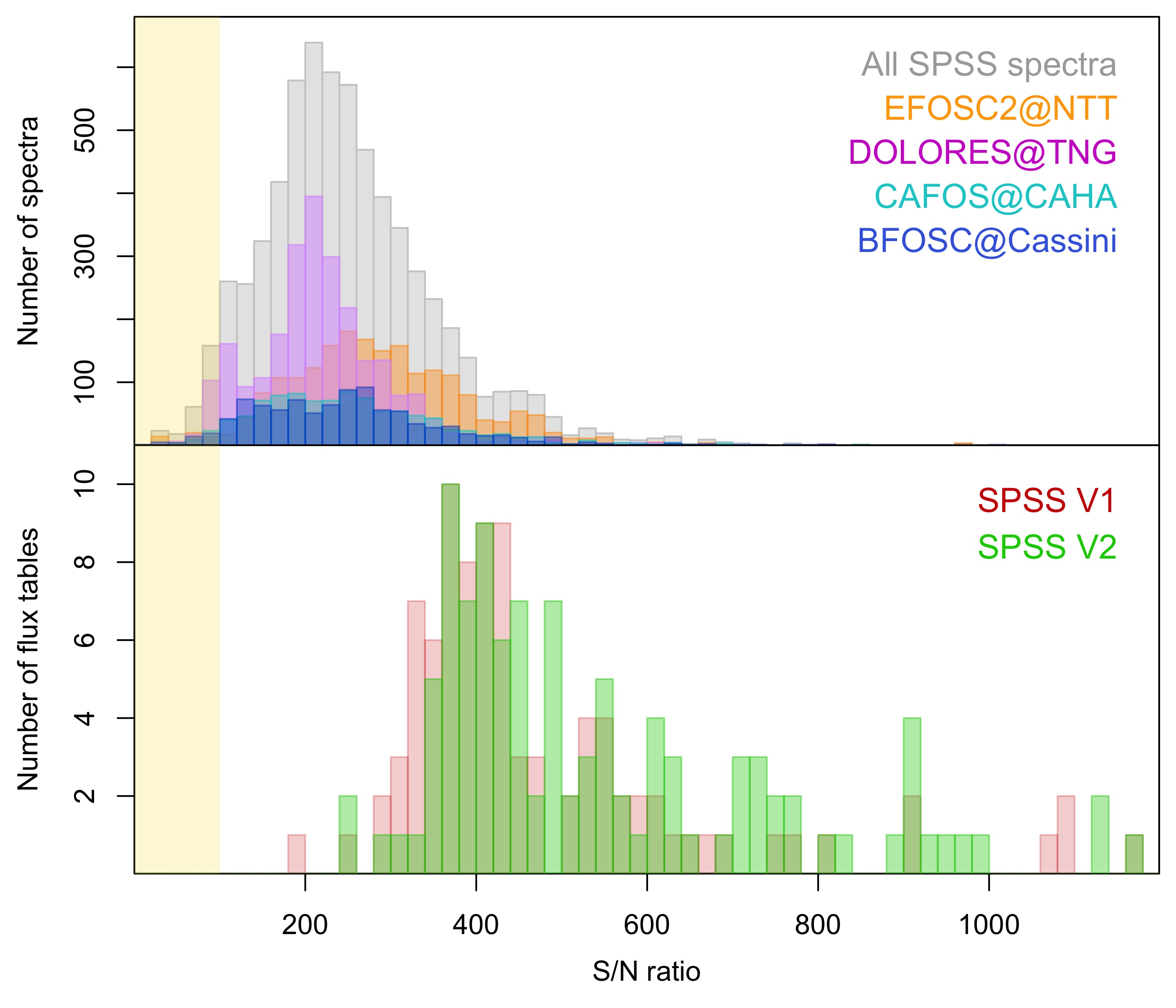} 
\vspace{-0.3cm}
\caption{Histogram of the S/N ratio measured on the central third of the individual spectra and SPSS final flux tables. {\em Top panel.} The entire set of 6466 good spectra is plotted in grey in the background, while the sets obtained with each facility are plotted in different colors, as labelled. {\em Bottom panel.} The flux tables in SPSS V1 (94 stars) and SPSS V2 (112 stars) are plotted in different colors, as labelled. A few flux tables with S/N ratio between 1200 and 2300 were left out to make the plot more readable. The region with S/N ratio$<$100, our S/N rejection criterium, is shaded in yellow in both panels.}
\label{fig:snr}   
\end{figure}

Once the 2D spectroscopic frames were pre-reduced as described above, we proceeded to extract the spectra and to calibrate them in wavelength using the \textsc{iraf} task apall \citep{valdes92}. To avoid flux losses, we observed the spectra with wide slits (10" or 12", depending on the spectrograph). A spectrum was considered {\em wide-slit} if the ratio between the slit and the seeing during the observations was higher than 6, while it was considered {\em narrow-slit} if it was between 1.5 and 6, and it was only used for wavelength calibration purposes if it was below 1.5. To extract the spectra, after tracing them in the direction of dispersion and subtracting the sky level, we summed all the pixels along the slit in a window as large as 6 times the FWHM of the spectrum. We did not use the optimal extraction option \citep{valdes92} but we strictly summed up the pixels to minimize flux losses, profiting from the fact that the S/N ratio of our spectra was always above 100 (Figure~\ref{fig:snr}). The sigma spectrum (the square root of the counts at each wavelength) was computed to represent the actual uncertainty. The sigma spectrum was processed consistently and carried along with the spectra in every step of the reduction (see also the next sections), so that the uncertainty associated to each reduction step was taken into account. 

To facilitate the wavelength calibration, that is complicated in the case of wide-slit observations, we also acquired narrow-slit spectra, generally using 2" or 2.5" slits. The narrow-slit spectra always preceded or followed immediately a wavelength calibration lamp, to minimize lamp flexure effects \citep{altavilla15}. In substance, we extracted the wavelength calibration lamps and used them to obtain a dispersion solution that was applied to both the narrow and wide spectra. The wide spectra were observed with long exposure times and repeated three times, thus they were generally observed far from the lamps. Moreover, the star centering within the slit was less controllable than with the narrow spectra. Therefore the wavelength calibration of wide spectra was adjusted by cross-correlation with the narrow spectra. With this procedure, and by comparing several spectra of the same star at the end, before creating the final flux tables (Section~\ref{sec:calib}) we achieved a wavelength calibration accuracy of 0.15-0.40~nm, with a median of 0.2~nm.

Each data frame was subject to a rigorous quality control (QC) procedure \citep{altavilla15} concerning the S/N ratio, the amount of saturated pixels, the presence of CCD defective pixels, the presence of artifacts or other problems. The QC system allowed for three levels: frames with no problems or with warnings were carried out to the following steps, while frames with serious problems were discarded. 

\begin{figure} 
\includegraphics[clip,width=\columnwidth]{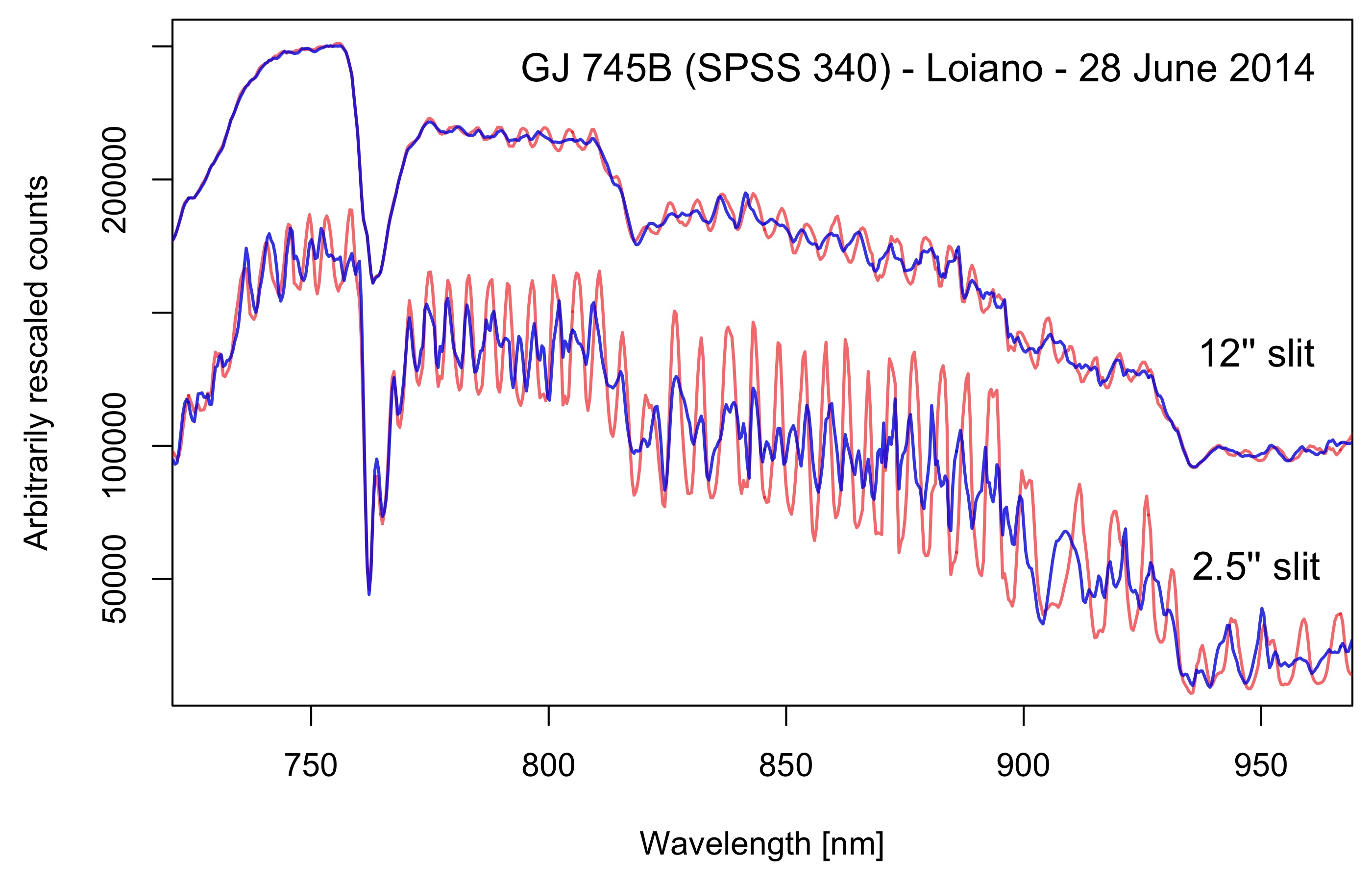} 
\vspace{-0.3cm}
\caption{Example of fringing mitigation for the Secondary SPSS GJ\,745B (SPSS\,340), observed repeatedly with BFOSC in Loiano on 28 June 2014. This is the facility with the strongest fringing pattern. Two of the seven uncorrected spectra are plotted in red, the corresponding corrected spectra are plotted in blue. The spectra taken with a 2.5'' and a 12'' slit are arbitrarily rescaled in flux for clarity, as labelled.}
\label{fig:fringe}   
\end{figure}

\subsection{Advanced spectra reductions}
\label{sec:reds}

The extracted spectra were further manipulated to mitigate some typical instrumental effects: fringing, flux losses in case of narrow-slit spectra, telluric absorption, and second-order contamination. 
In the case of fringing, no procedure can fully remove its signature from spectra, unlike in the imaging case. Fringing only affected our red grisms and appeared much more pronounced in the case of narrow-slit spectra (see Figure~\ref{fig:fringe}). We adopted the mitigation procedure described in details by \citet{altavilla15}, that is based on the method originally adopted for STIS spectra \citep{malumuth03}. Briefly, the fringing pattern present in the spectroscopic flat fields is shifted and scaled and then applied to the spectra, to minimize the fringing oscillations. The procedure is much more effective when the fringing pattern is more pronounced (see Figure~\ref{fig:fringe}). The typical strength of the fringing pattern and the approximate starting wavelength for each facility can be found in Table~4 by \citet{altavilla15}. In summary, fringing starts at about 700--800~nm and varies from negligible to 15\%. With our procedure, we could beat down fringing to below 3\% in most spectra, except for a minority of cases. This uncertainty was used to update the sigma spectrum in the relevant wavelength regions. When combining different spectra for a given SPSS, fringing was further reduced to about 1\% or less, except for a few SPSS, because the fringing patterns are rarely aligned in wavelength for spectra obtained in different nights or with different facilities. Note that the fringing mitigation procedure was applied to the SPSS V2 spectra but not to the SPSS V1 ones\footnote{The SPSS V1 release was used to calibrate {\em Gaia} DR2, which only contained integrated photometry, and thus the effect of fringing was expected to be negligible. In DR3 however, the first BP and RP spectra will be released. The fringing oscillations can create problems in the external calibration, at least locally, thus fringing mitigation was applied to V2 spectra.}.

\begin{figure} 
\includegraphics[clip,width=\columnwidth]{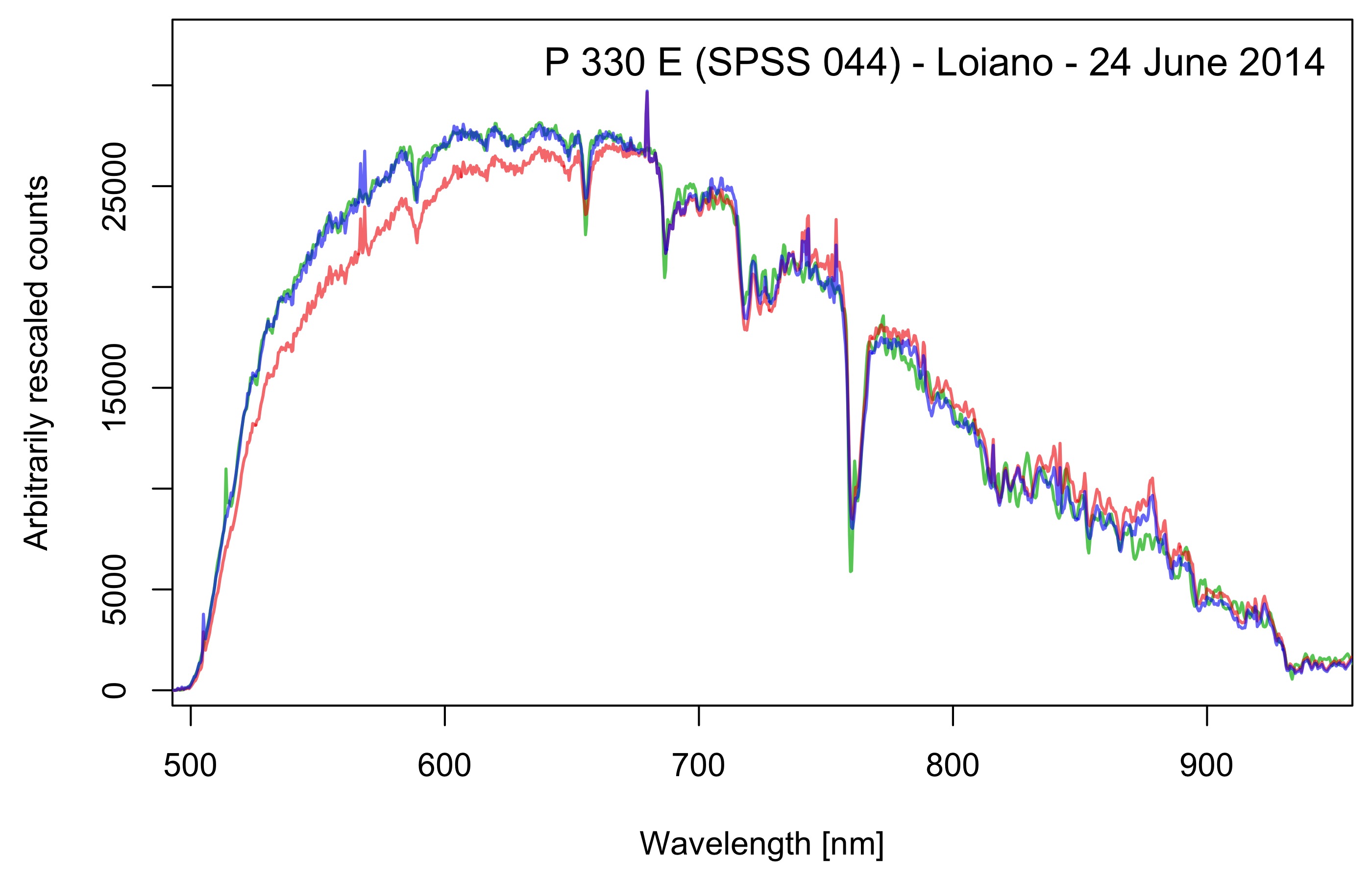} 
\vspace{-0.3cm}
\caption{Example of slit-loss correction for a narrow-slit spectrum of the Primary SPSS P\,330\,E (SPSS\,044), observed with BFOSC on the 24 June 2014. The reference wide-slit spectrum is plotted in green, the uncorrected narrow-slit spectrum in red, and the corrected one in blue.}
\label{fig:slitloss}   
\end{figure}

As mentioned, all the narrow-slit spectra, with a slit width smaller than 6 times the seeing and larger than 1.5, were later used together with the wide spectra to create the final flux tables (see Section~\ref{sec:abscal}). They were corrected for wavelength dependent slit losses, caused by the fact that the seeing is larger at shorter wavelength and thus the flux loss is generally larger in the blue. Spectra with slits narrower than 1.5 times the seeing were only used for the wavelength calibration and then discarded. The actual slit loss at each wavelength can be complicated by other effects such as atmospheric diffraction\footnote{To minimize atmospheric diffraction effects, we always aligned the slit with the star's parallactic angle. Exposures were rarely longer than one hour and we did not observe the SPSS too close to the horizon.} and misalignment of the star in the slit, thus it cannot be modelled from first principles. The procedure relied on a wide-slit spectrum of the same SPSS, possibly obtained in the same night. The ratio between the narrow-slit and the wide-slit spectrum was fitted with a smooth function (typically a cubic spline of low order). The function was then applied to the narrow spectrum to partially recover the lost flux. The sigma spectra were worsened according to the correction applied.  

\begin{figure} 
\includegraphics[clip,width=\columnwidth]{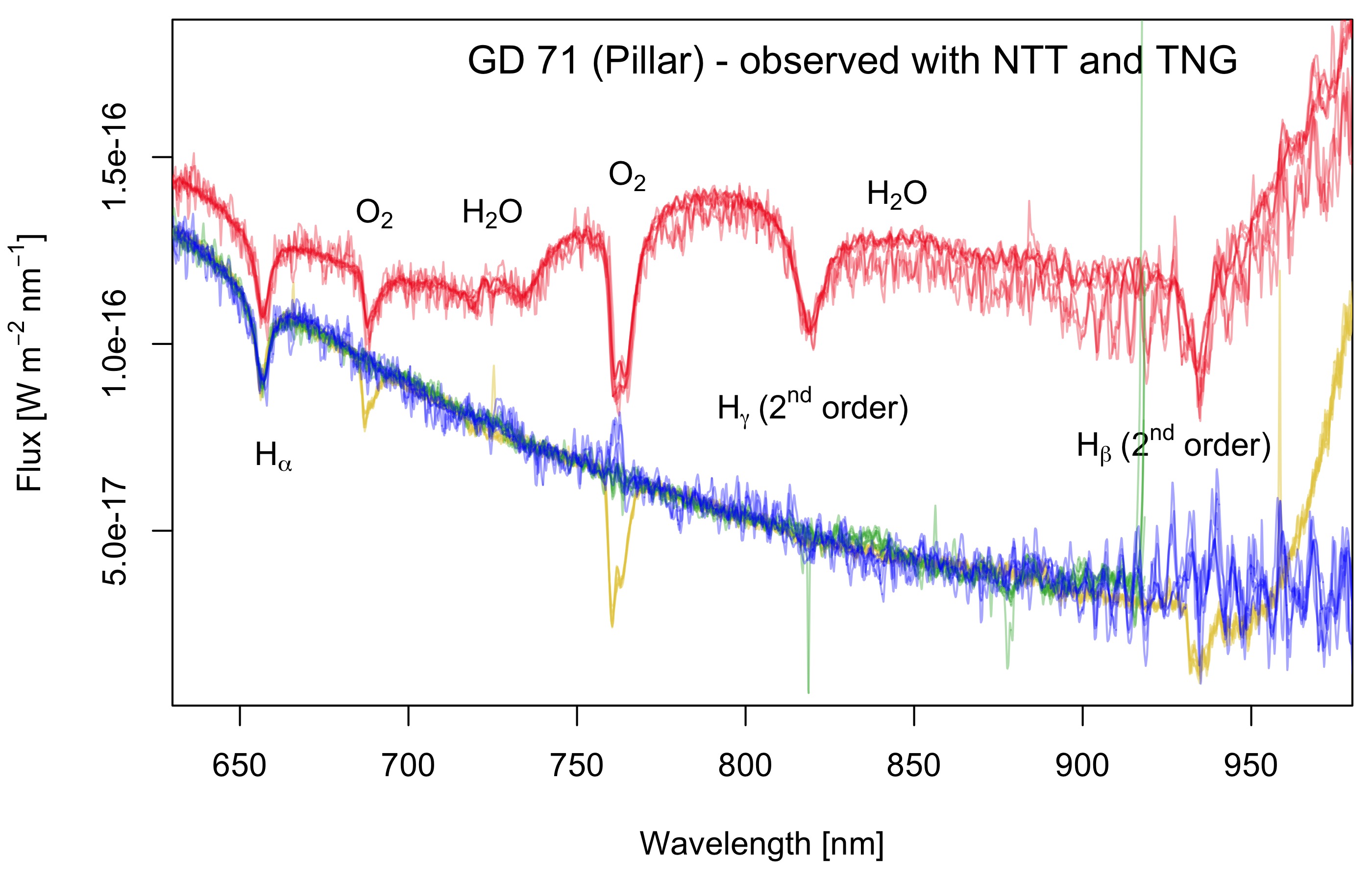} 
\vspace{-0.3cm}
\caption{Example of second-order correction for the Pillar GD\,71 (SPSS 002). The V2 flux table in the red wavelength range is based on 24 wide-slit spectra observed with TNG and NTT. Of these, the 12 NTT spectra observed with grism \#16, affected by a significant second-order contamination, are plotted in red. The corresponding second-order corrected spectra are plotted in blue. The uncorrected TNG spectra, which are unaffected up to $\simeq$950~nm, are plotted in gold. The NTT spectra obtained with grism \#5, unaffected but reaching only up to $\simeq$920~nm, are plotted in green.}
\label{fig:2ndorder}   
\end{figure}

\begin{figure} 
\includegraphics[clip,width=\columnwidth]{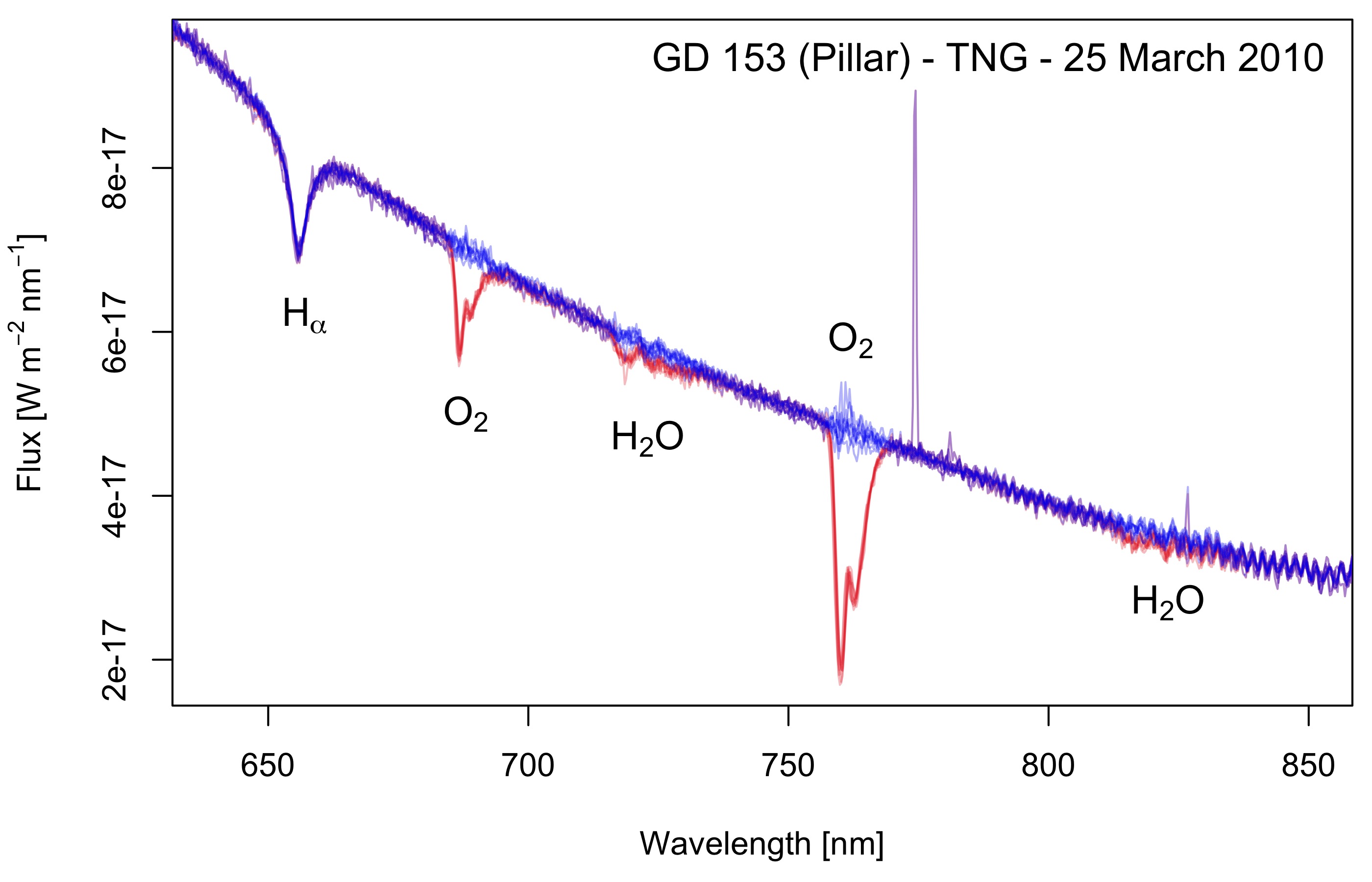} 
\vspace{-0.3cm}
\caption{Example of telluric correction for the Pillar GD\,153 (SPSS\,003), observed repeatedly with DoLoRes at the TNG on 25 March 2010. Nine uncorrected spectra are plotted in red, the corresponding nine corrected spectra are plotted in blue. The H$_{\alpha}$ line is marked as well as the main telluric absorption bands in the plotted wavelength range.}
\label{fig:tell}   
\end{figure}

We then took care of second-order contamination. With the goal of covering the {\em Gaia} wavelength range with two grisms at most, at the desired resolution, we had to use grisms affected by second-order contamination, i.e., light from the blue side of the second order contaminating the red side of the first order spectra. This occurred for two of the four spectrographs: EFOSC2 and DoLoRes. In the case of DoLoRes, only the reddest 50~nm were affected in the LR-R grism (see Table~\ref{tab:tels} and Figure~\ref{fig:2ndorder}). In the case of EFOSC2, however, almost the entire range of grism \#16 was heavily contaminated, so we added some exposure with grism \#5 (Table~\ref{tab:tels}) that covered a smaller wavelength range, but was free from second-order contamination. For the same reason, we observed each SPSS with more than one facility whenever possible. We used the grism \#5 spectra to validate our second-order correction procedure. The procedure is described in details in \citet{altavilla15} and is based on a method originally proposed by \cite{sanchez06}. An example of the correction can be found in Figure~\ref{fig:2ndorder}. The typical residuals of the correction are of 1--2\% in flux on the single spectrum and well below 1\% on the final flux tables. The sigma spectrum of each affected star was worsened accordingly to take into account the related uncertainty.

Finally, the removal of telluric absorption features was done using two different methods. The SPSS V1 spectra were corrected with our in-house procedure based on the \textsc{iraf} telluric task. We used theoretical atmospheric absorption spectra from the HITRAN \citep{hitran} or SPECTRA \citep{spectra} databases\footnote{Accessed through: \url{http://spectra.iao.ru/}} for H$_2$O molecules and a library of our observed SPSS spectra for O$_2$ ones, and explored the libraries until the residuals of the correction were minimized. New spectra added in the SPSS V2 version were instead corrected using Molecfit, which is based on similar data souces to build theoretical atmospheric absorption spectra \citep{molecfit,molecfit2}. Molecfit is more practical to use and we verified that the two methods produce comparable results. The typical residuals on the single spectrum after the telluric correction were smaller than 2\% in the weakest telluric bands, and could reach up to 5\% or more in the strongest ones. However, in the final merging of all the spectra for each given SPSS, the telluric residuals were reduced to less than 1--2\% in all but a handful of cases (see Figure~\ref{fig:tell} for an example). The sigma spectrum was thus worsened in the telluric band regions accordingly.

\section{SPSS flux calibration}
\label{sec:calib}

The SPSS reference system is tied to the CALSPEC reference system \citep{bohlin14,bohlin19}, because we used as {\em Pillars} of our flux calibration \citep[see][for details]{pancino12} the three pure hydrogen white dwarfs (WD) G191-B2B, GD 71, and GD 153 \citep{bohlin95}, which are directly calibrated on Vega\footnote{A fourth star, HZ\,43, was originally considered but later discarded because of a companion at 3" \citep{bohlin01}, that could disturb ground-based observations.} and on Sirius \citep{bohlin14}. More precisely, the SPSS V1 and SPSS V2 versions are calibrated on the CALSPEC system as it was in 2013 (see also Section~\ref{sec:calspec}). In practice, we observed at least one of the three CALSPEC pillars each night. When they were not visible, especially from the Southern hemisphere, we observed a list of well-behaved and well characterized stars from the CALSPEC database and from the literature as our calibrators, the so called {\em Primary} SPSS \citep[see][for details]{pancino12}. The reference flux tables used to calibrate the SPSS V1 and V2 sets are listed in Table~\ref{tab:calibs}.

\begin{figure} 
\includegraphics[clip,width=\columnwidth]{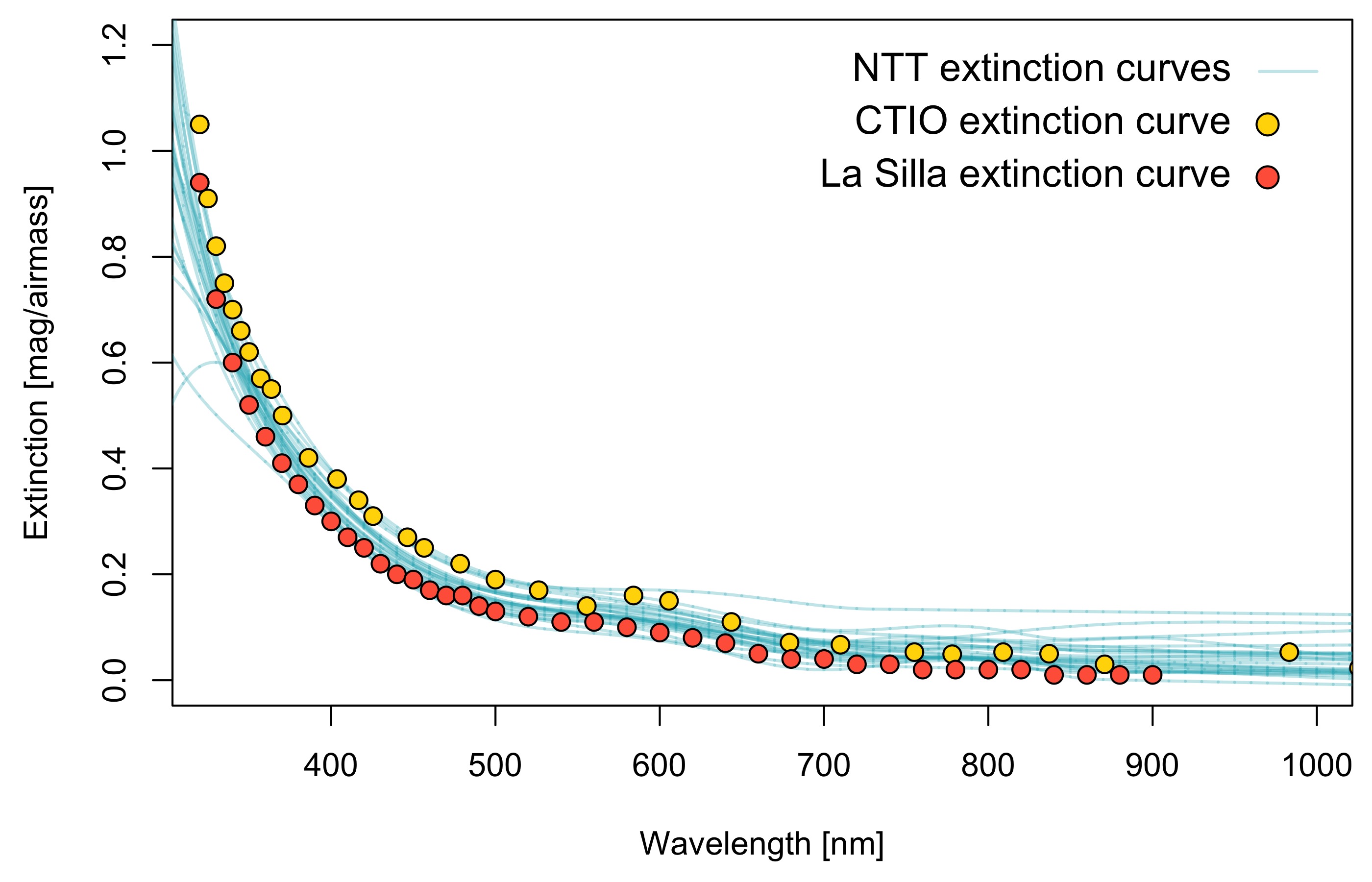} 
\vspace{-0.3cm}
\caption{The extinction curves obtained during 25 NTT observing nights with varying atmospheric conditions (cyan lines), compared with the Cerro Tololo (CTIO) reference curve \citep[][yellow points]{stone83,baldwin84} and with the ESO La Silla extinction curve \citep[][red points]{eso93}.}
\label{fig:ext}   
\end{figure}
\begin{figure} 
\includegraphics[clip,width=\columnwidth]{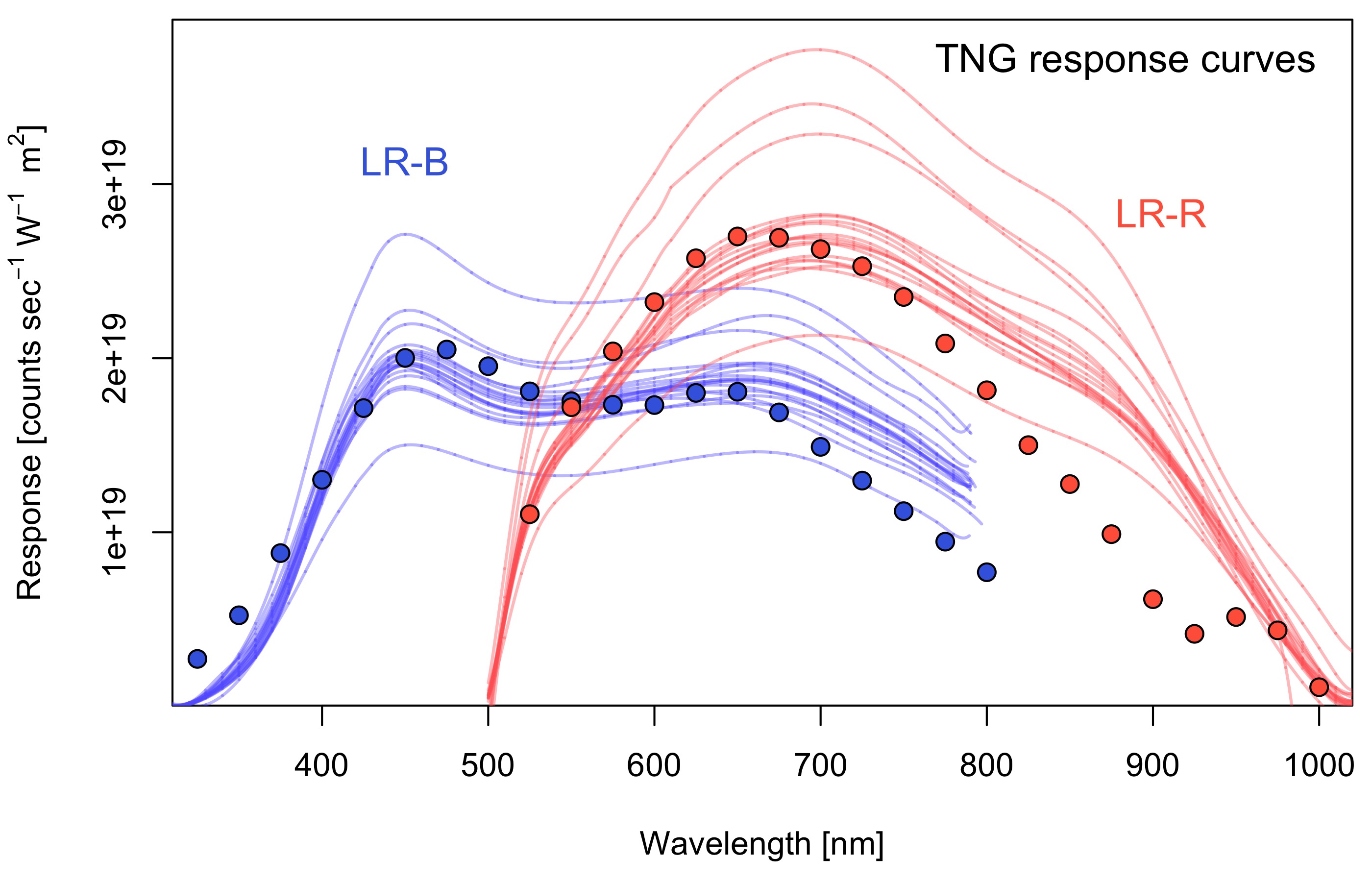} 
\vspace{-0.3cm}
\caption{The red and blue response curves (red and blue lines) obtained during 19 TNG observing nights, from 2008 to 2010, with varying atmospheric conditions. The LR-R and LR-B curves obtained by the TNG staff in 2005 (with a different CCD) are shown for reference (red and blue dots, rescaled).}
\label{fig:resp}   
\end{figure}

\subsection{Night calibration and quality assessment}
\label{sec:relcal}

To correct for the atmospheric extinction, we computed extinction curves, employing all the observations of stars that were repeated with a minimum time difference of 1~hrs and an airmass difference of at least 0.2. We monitored the nightly stability of the curves and we divided the nights in three categories: {\em photometric} when there were no significant slopes or shape variations among the different curves obtained during the night and all the curves were within 1\% from each other\footnote{Given the quality degradation at the spectra blue and red borders, the QC in this and in the following calibration steps was performed always in the 400--800~nm range.}; {\em grey} when there were no slope or shape variations, but the curves differed from each other by more than 1\%; and {\em non-photometric} when there were slopes or shape variations, or when the curves differed by more than 3\%, or when the available data did not span the minimum time and airmass ranges of 4~h and 0.2, respectively. The final extinction curves were computed as the median of the curves obtained for the night and an example is shown in Figure~\ref{fig:ext}, where we compare our NTT curves, obtained in varying atmospheric conditions, with literature curves for La Silla and Cerro Tololo. The comparison is satisfactory within 400--800~nm approximately, but there is still room for improvement at the two extremes of the wavelength range. 

\begin{table}
\caption{List of the reference flux tables of Pillars and Primary SPSS used to calibrate the V1 and V2 releases.}\label{tab:calibs}
\begin{tabular}{|l|l|l|l||}
\hline
SPSS & Name & Source & Version \\
\hline                                                    001 & G\,191-B2B       & CALSPEC & g191b2b\_mod\_007.fits\\
002 & GD\,71     & CALSPEC & gd71\_mod\_007.fits \\
003 & GD\,153    & CALSPEC & gd153\_mod\_007.fits \\
007 & HZ\,2      & CALSPEC & hz2\_005.fits \\
009 & LTT\,2415  & \multicolumn{2}{l}{\citet{hamuy94}} \\
010 & GD\,108    & CALSPEC & gd108\_005.fits \\
011 & Feige\,34  & CALSPEC & feige34\_stis\_001.fits \\
013 & Feige\,66  & CALSPEC & feige66\_002.fits \\
015 & GRW+705824 & CALSPEC & grw\_70d5824\_stisnic\_003.fits \\
018$^*$ & BD+284211 & CALSPEC & bd\_28d4211\_stis\_001.fits \\
020$^*$ & BD+174708 & CALSPEC & bd\_17d4708\_stisnic\_003.fits \\
023 & Feige\,110 & CALSPEC & feige110\_stisnic\_003.fits \\
036 & 1805292    & CALSPEC & 1805292\_nic\_002.fits \\
037 & 1812095    & CALSPEC & 1812095\_nic\_002.fits \\
043 & P\,177 D   & CALSPEC & p177d\_stisnic\_003.fits \\
046 & KF\,06\,T1 & CALSPEC & kf06t1\_nic\_001.fits \\
\hline
\end{tabular}\\
$^*$These two Primary SPSS were used as calibrators in V1. Later, they were found to be likely variable by \citet{marinoni16}. Therefore, in V2 we did not use spectra from nights calibrated with these SPSS to build the median reference spectrum of each SPSS (see Section~\ref{sec:calib}). However, once re-calibrated, we used the spectra to build the final flux tables.
\end{table}

We used the spectra obtained for the night calibrators, either Pillars or Primary SPSS (Table~\ref{tab:calibs}), to compute the response curves for the night calibration. We generally observed from one to three different calibrators, at different airmass during the night. The night response curve was obtained as the median of the various curves obtained from each observation of the calibrators and the uncertainty was obtained using the median absolute deviation (MAD). We note here that the final SPSS version will be calibrated in a two-step procedure, in which the Primary SPSS will be calibrated on the Pillars before being used to calibrate all the other (i.e., the {\em Secondary}) SPSS \citep{pancino12}. In the V1 and V2 versions of the SPSS library presented here, we calibrated each night directly on the best available night calibrators, thus the internal consistency is expected to improve slightly in the next SPSS release. For the calibration presented here, we used the literature reference flux tables listed in Table~\ref{tab:calibs} to compute the response curves. The curves obtained for TNG in several observing nights with different atmospheric conditions are shown in Figure~\ref{fig:resp}. 

After the spectra observed in each night were calibrated with the extinction and response curves computed as described above, we further checked that repeated wide-slit exposures of the same SPSS during the night were compatible with each other within 1\% and did not display any significant shape variation. A warning was issued in case of zero-point variations up to 3\%, and the spectra were discarded in case of higher zero-point variations and/or significant shape variations. 

\subsection{Absolute calibration and flux tables creation}
\label{sec:abscal}

To select the spectra entering the two preliminary SPSS V1 and V2 releases, we used different criteria. For V1, we selected all the nights that were judged to be photometric according to the criteria described in the previous section. When V1 was internally released, the spectrophotometric campaign was not completed yet, thus we could count on 15 observing nights. For V2, which occurred after the end of the observing campaigns, we could count on 26 nights, using the same criterium. As an exception, we included a few nights that were judged {\em grey} or even {\em non photometric} to include a few more red stars, that were necessary to better keep under control the calibration performance at extreme colors for the {\em Gaia} DR2 and (E)DR3 data\footnote{The actual {\em Gaia} external calibration procedure is quite complex and the details can be found in \citet{evans18} for DR2, in \citet{riello20} for EDR3, and in the forthcoming DR3 release papers (De Angeli et al. and Montegriffo et al., in preparation).}. The final flux tables of the few SPSS that did not contain at least one observation sequence in a photometric night were issued with a warning.

Once the data were selected, we compared all the available spectra for each SPSS with each other. We first selected only the wide-slit spectra of a given SPSS that were observed in photometric nights and rejected those that were not within 1\% from the median spectrum, paying special attention to the region of overlap of the red and blue grisms. We then computed the median spectrum with its median absolute deviation (MAD). In doing so, we slightly adjusted the wavelength calibration of all spectra to align them to the median spectrum, using cross-correlation techniques. 
We then used the median of the best spectra to adjust the flux calibration of all the remaining spectra. In this way, only the best photometric data contributed to the {\em accuracy} of the final flux table, while all the spectra contributed to smooth out the pixel-to-pixel variations caused by noise and by the residuals of the telluric, fringing, and second-order corrections (when relevant). At this point, we recomputed the final median spectrum using all the spectra, the best ones plus all the re-calibrated remaining ones, to create the final SPSS flux table. In a few cases, we observed that some spectra showed a significant slope when compared to the median of the photometric spectra: we included them only when the slope could be successfully removed (within 1\% of maximum deviation). In future SPSS releases we will use our absolutely calibrated photometry \citep{altavilla20} to validate and -- if necessary and appropriate -- to re-adjust the  calibration of spectra taken in non perfectly photometric conditions (see also Section~\ref{sec:v3}).

\begin{table*}
\caption{Parameters of the best-fitting models used to extend the SPSS flux tables to cover the full  300--1100~nm {\em Gaia} range. Stars extended with observed spectra are not listed here. Only the first five rows are shown, the table is available in its entirety online (see Data Availability Section).}\label{tab:ext}
\begin{tabular}{|l|l|r|r|r|r|r|l|l|}
\hline
ID & Name & SpType & T$_{\rm{eff}}$ & log\,$g$ & [Fe/H] & A$_{\rm V}$ & Library & Notes \\
& & & (K) & (dex) & (dex) & (mag) \\
\hline 
005	& EG\,21 & DA3 & 16000 & 8.0 & 0.0 & 0.0 & \citet{koester10}	& Template agreement at junctions 1--3\% \\
006 & GD\,50 & DA2 & 43000 & 8.8 & 0.0 & 0.0 & \citet{sordo11} & Template agreement at junctions 1--3\% \\
008 & LTT\,3218	& DA & 9250 & 7.75 & 0.0 & 0.0 & \citet{koester10}	& --- \\
009 & LTT\,2415 & G	& 23900	& 7.0 &	0.0 & 0.045 & \citet{sordo11} & Discrepant spectral type \\
010	& GD\,108 & B & 26030 & 4.675 & 0.0 & 0.1210 & \citet{sordo11} &  Template agreement at junctions $>$3\% \\
\hline
\end{tabular}\\
\end{table*}

\begin{figure} 
\includegraphics[clip,width=\columnwidth]{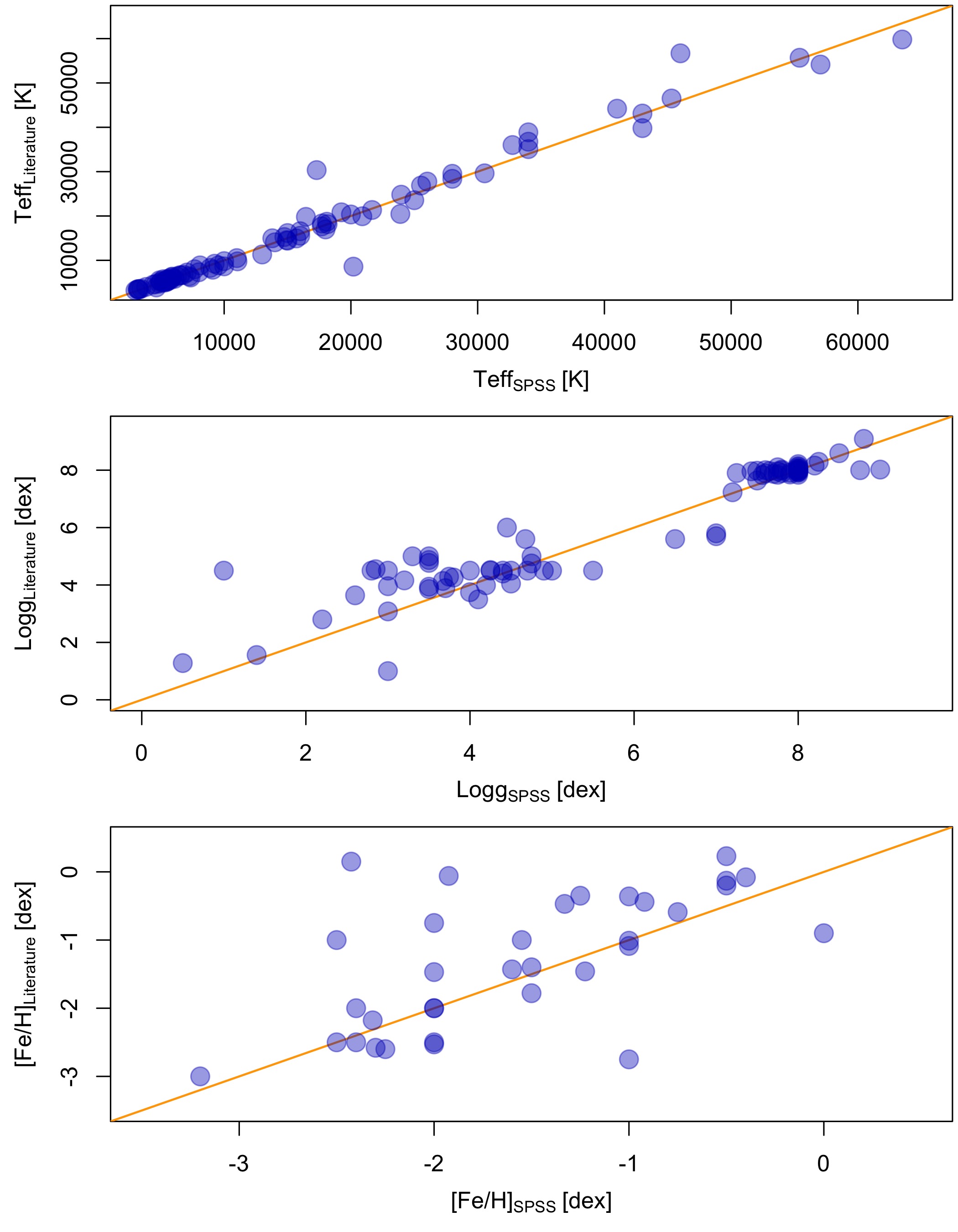} 
\vspace{-0.3cm}
\caption{Comparison of the parameters of the best-fitting template spectra (abscissae in all panels) with the available literature determinations (ordinates). The top panel shows the effective temperature, the middle one the surface gravity, and the bottom one the metallicity. The 1:1 line is plotted in orange in all panels (see Section~\ref{sec:mod} for more details).}
\label{fig:mod}   
\end{figure}

\subsection{Extension with template spectra}
\label{sec:mod}

As mentioned, the sensitivity of the used telescope and instrument combinations drops at the borders of the spectrum. Most of the used grisms do not reach 1050~nm (Table~\ref{tab:tels}) and when they do the S/N ratio is much lower than in the central part of the spectrum. Similar considerations apply to the blue side. For this reason, all the QC and validation procedures are limited to the 400--800~nm range. Some of these limitations are mitigated when combining several different spectra to obtain an SPSS flux table. Nevertheless, it was necessary to extend the spectra to cover the full nominal {\em Gaia} wavelength range: the passbands cover the 330--1050~nm range approximately, but the {\em Gaia} BP and RP spectra formally start at 300~nm and reach 1200~nm, although there is very little flux before 330~nm or after 1050~nm. Therefore, we extended the SPSS flux tables to cover the 300--1100~nm range. We note  that {\em Gaia} does have limited sensitivity at the extremes of the spectral range, thus any imperfection in the spectra extension of a given SPSS will have a limited impact on the final accuracy of the {\em Gaia} flux calibration. 


We followed a procedure similar to the one adopted by the CALSPEC team: we extended our observed spectra using template specta (either observed spectra or theoretical models). In our case however, we did not use the templates to calibrate the spectra, but we adjusted the templates onto the absolutely calibrated spectra. For the stars in common with the CALSPEC library, we extended the SPSS spectra by adjusting the corresponding observed CALSPEC spectra ($\lesssim$1\% adjustments). For the other SPSS, we used a series of different observed, theoretical, or semi-empirical libraries\footnote{Most of the spectra were retrieved from the collection at \url{http://svo2.cab.inta-csic.es/theory/}.} depending on the spectral type of the SPSS \citep{stritzinger05,palacios10,koester10,sordo11,husser13,apsis,coelho14,levenhagen17}.
We used a $\chi^2$ minimization algorithm to chose the best-fitting model for each SPSS. During the comparison, the template spectra were aligned in wavelength onto the observed ones with a cross-correlation. Whenever literature parameters were available, we used them as a first guess to reduce the parameter space to explore, otherwise we used the literature spectral type (Table~\ref{tab:spsslist}). We found a relatively good agreement between our best-fitting parameters and the literature ones, except for metallicity which is typically poorly constrained for hot stars. The typical median differences and MADs we found are $\Delta$T$_{\rm{eff}}$=--60$\pm$697~K, $\Delta$log$g$=--0.14$\pm$0.33~dex, and $\Delta$[Fe/H]=--0.17$\pm$0.56~dex (see also Figure~\ref{fig:mod}). The results of the procedure for the stars that were extended with theoretical models can be found in Table~\ref{tab:ext}, where we list the best-fitting parameters along with the source of the theoretical spectra. 

\begin{figure} 
\includegraphics[clip,width=\columnwidth]{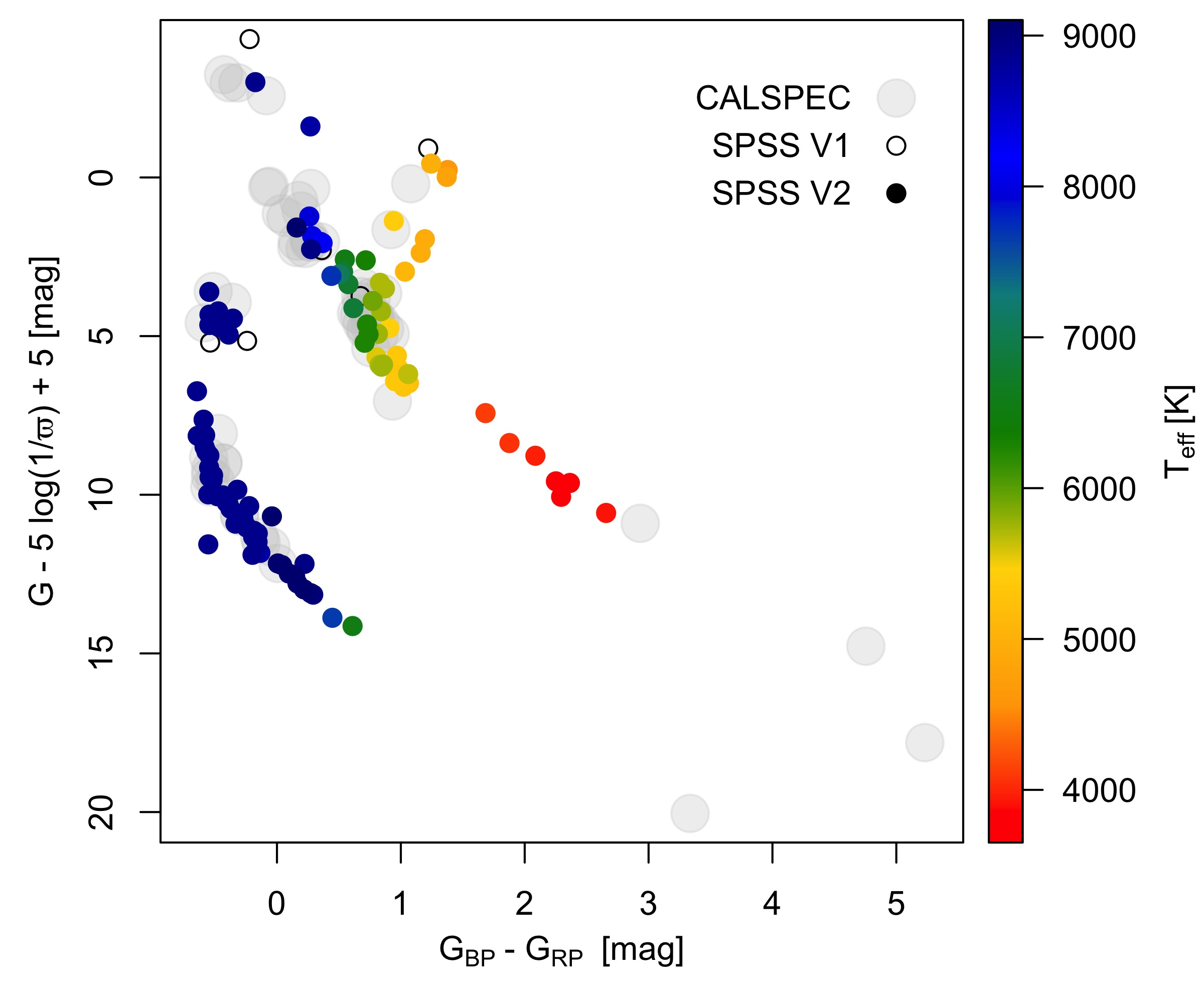} 
\vspace{-0.3cm}
\caption{The {\em Gaia} colour -magnitude diagram of the SPSS. The CALSPEC stars are plotted in the background as grey circles for comparison. The SPSS V2 stars are colored as a function of their T$_{\rm{eff}}$ (the color scale saturates at T$_{\rm{eff}}>9000$~K, for readability). Seven SPSS V1 stars that were not included in SPSS V2 are plotted as empty black circles.}
\label{fig:cmd}   
\end{figure}

\section{SPSS results and validation}
\label{sec:res}

The final SPSS V1 and V2 libraries contain respectively 94 and 112 stars. The flux tables are available online (see Data Availability Section) as ascii files containing the wavelength resampled to 0.1~nm and the flux with its uncertainty expressed in W~m$^{-2}$ nm$^{-1}$. Seven of the SPSS in V1 were later rejected because of variability \citep{marinoni16}, companions, or problems with the spectra, and thus do not appear in V2. On the other hand, 25 new SPSS were added in V2. As a result, V2 contains more SPSS than V1 and on average more spectra are used to create the flux tables of each SPSS than in V1 (see also Figure~\ref{fig:snr}). In any case, the two versions of the SPSS library are on the same spectro-photometric flux scale within $\simeq$0.1--0.2\%, as detailed in Table~\ref{tab:calspec}. 

The current list of SPSS V2 surviving candidates, along with the relevant information, can be found in Table~\ref{tab:spsslist}. Figure~\ref{fig:cmd} shows the color-magnitude diagram of the SPSS V1 and V2 sets, and is built using {\em Gaia} EDR3 parallaxes and photometry. As discussed in Section~\ref{sec:sele}, the SPSS grid does not contain extreme spectral types: only a handful of bright and blue stars, carefully selected to be constant and single, are included, while no star cooler than about 3000~K is included. However, compared to the current CALSPEC selection, the SPSS sample extends the WD sequence significantly towards the cool end, it fills the CALSPEC gap in the M dwarfs (1$\lesssim$G$_{\rm{BP}}$--G$_{\rm{RP}}\lesssim$3~mag), and it considerably increases the samples of hot subdwarfs and FGK dwarfs and giants. In the following, we present some validation tests on the SPSS V2 sample.

\begin{figure} 
\centering
\vspace{-0.3cm}
\includegraphics[clip,width=1.05\columnwidth]{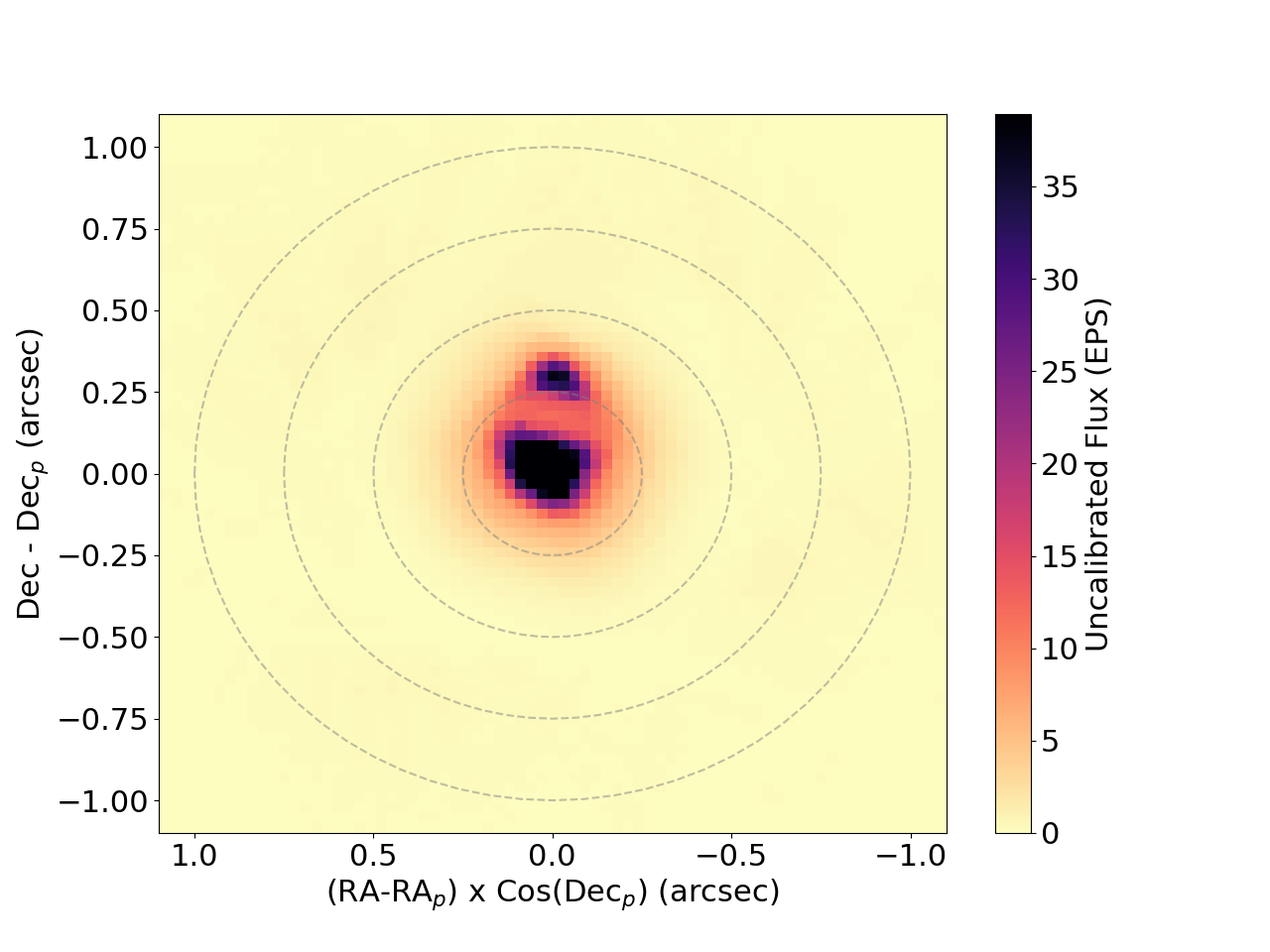} 
\includegraphics[clip,width=1.05\columnwidth]{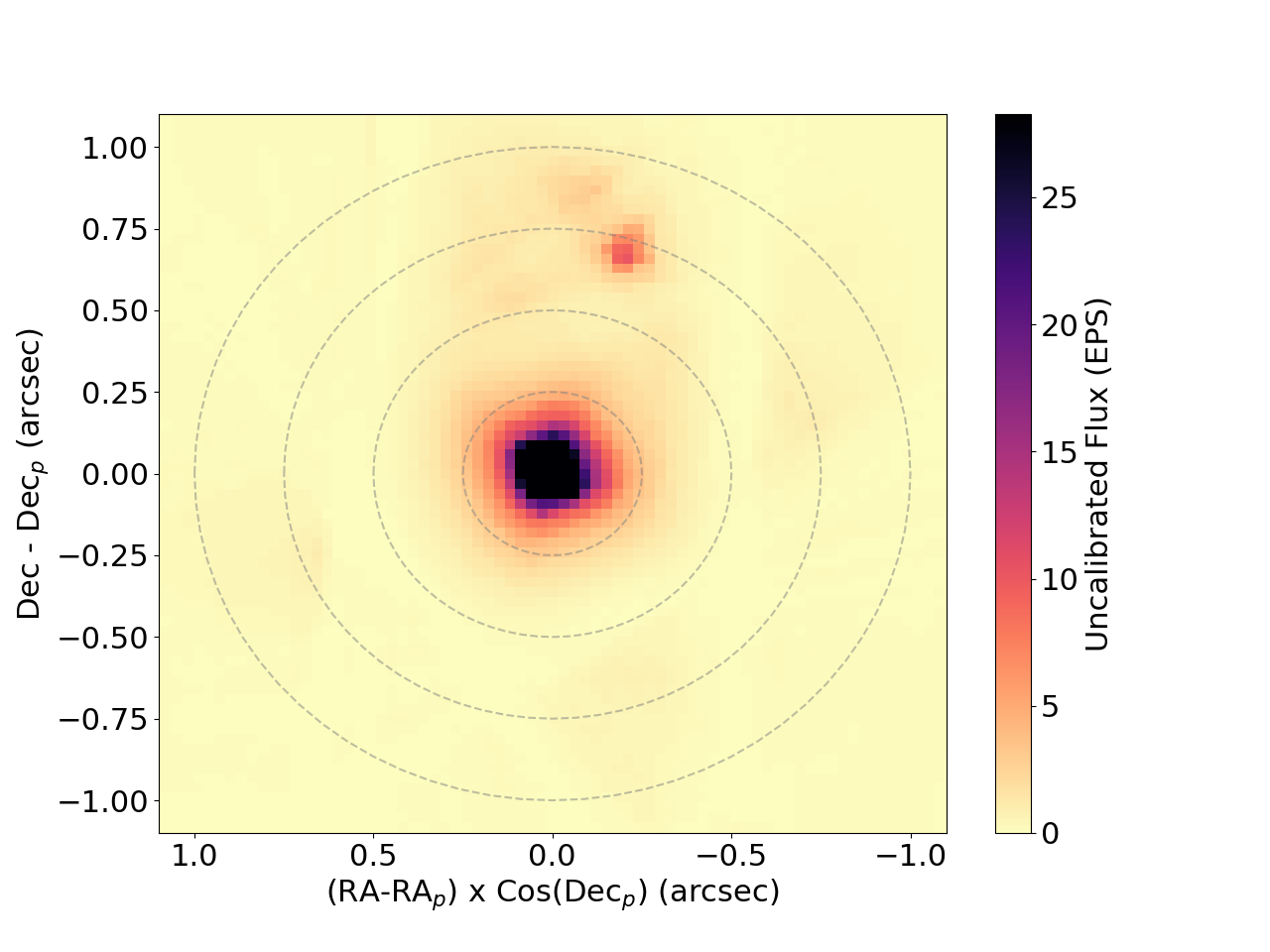}
\vspace{-0.3cm}
\caption{Images reconstructed as part of the SEAPipe analysis for two SPSS sources found to have visible companions. Images are reconstructed centred on the location of the primary source ($\mathrm{RA}_p,\mathrm{Dec}_p$), where the primary source refers to the {\em Gaia} source whose transit data is used in the reconstruction.
The grey dashed circles indicate angular separations of 0.25, 0.50, 0.75, and 1.0" from the centre of the image (and primary source position).
The colour scale of these images was set ignoring the image pixels within 0.1" of the centre, in order to better visualise the companions, hence the saturated appearance of the primary sources.
{\em Top panel:} The reconstruction for SPSS\,192, showing a companion at an angular separation of $\simeq$0.29" and a magnitude difference of $\simeq$1.9~mag. This companion is not present in EDR3. {\em Bottom panel:} The reconstruction for SPSS\,215, showing the companion present in EDR3 at an angular separation of $\simeq$0.69" and a magnitude difference of $\simeq$3.0~mag. }
\label{fig:sea}   
\end{figure}


\subsection{SPSS list pruning with {\em Gaia}}
\label{sec:sea}

As mentioned, we carried out a short-term constancy monitoring (1--2~h) on all SPSS that needed to be tested \citep{marinoni16}. We also started a long-term monitoring campaign (1--3~yrs), but we could not conclude it for lack of resources. Up to {\em Gaia} DR2, we relied mostly on literature information to remove long-term variables from the SPSS list. However, {\em Gaia} is expected, especially in the forthcoming releases, to be extremely efficient in the detection of long-term flux variations above 1\% \citep{mowlavi18}. So far, we searched for our SPSS candidates in the {\em Gaia} DR2 variable stars catalogue \citep{holl18} and we did not find any. With future {\em Gaia} releases, we will update our search to exclude all known variable stars. We also looked into the main EDR3 catalogue for companions to our SPSS V2 stars and found that 10 indeed have one, either farther than $\simeq$2" (our typical seeing is $\simeq$1"), or fainter than 5~mags (i.e., less than 1\% flux contamination), or both. The only exception (but see below) is SPSS\,215, which appears to have a companion in the main {\em Gaia} EDR3 catalogue at about 0.7", with a magnitude difference of about $\Delta$G=3~mag. This means that the ground-based spectra are certainly disturbed by the companion. 


To explore the problem further, we used the Source Environment Analysis (SEAPipe), which had its first runs in the {\em Gaia} processing in preparation of DR3. The aim of SEAPipe is to combine the transit data for each source and to identify any additional sources in the local vicinity. 
The first operation in SEAPipe is image reconstruction, where a two-dimensional image is formed from the mostly one-dimensional transit data (G$>$13~mag). The algorithm used to perform the image reconstructions is described in \cite{harrison11}. 
As part of the validation of SEAPipe all images reconstructed for SPSS sources were visibly inspected. Not all SPSS sources had reconstructed images, as there must be a good coverage of transits crossing the source in different orientations for a successful image reconstruction, this in fact is the main reason why SEAPipe was not run in previous releases.
SEAPipe was attempted on all 119 SPSS sources (the union of V1 and V2) and successfully produced images for 116 of these sources. The remaining 3 sources (SPSS\,20, 28 and 32) 
did not have enough suitable transits for the image reconstruction to proceed.
The inspection of these 116 images revealed 114 SPSS sources with no evidence of a companion within 1" at least to the level of the noise in their reconstructed images.
However, it was discovered that two of the sources, SPSS\,192 and 215, had companions within 1". These images are shown in Figure~\ref{fig:sea}.
The angular separation (0.29", 0.69") and magnitude difference (1.9, 3.0~mag) of the companion from the SPSS source (192, 215) were found by performing an image parameter analysis on the data used to construct the images. For SPSS\, 215 these are in agreement with those of the companion found in EDR3.

We decided to leave SPSS\,192 and 215 in the list because they were actually used in the calibration procedure for EDR3, but they will be rejected from the next SPSS release. They were annotated in Table~\ref{tab:spsslist} and assigned a quality flag of two. 

\begin{table}
\caption{Median ratio and MAD of the SPSS V1 and V2 libraries and of the SPSS and the CALSPEC libraries (2021 version), for the 11 stars in common (see Figure~\ref{fig:calspec}).}\label{tab:calspec}
\begin{tabular}{|l|r|r|r|r|}
\hline
Ratio & 300--400~nm & 400--800~nm & 800--1070~nm \\
\hline 
V2/V1 & 1.0010$\pm$0.0025 & 1.0001$\pm$0.0011 &
1.0018$\pm$0.0034 \\
V1/CALSPEC & 0.9921$\pm$0.0207 & 1.0016$\pm$0.0056 & 1.0012$\pm$0.0152 \\
V2/CALSPEC & 0.9891$\pm$0.0101 & 0.9995$\pm$0.0089 & 1.0052$\pm$0.0115 \\
\hline
\end{tabular}
\end{table}

\begin{figure} 
\includegraphics[clip,width=\columnwidth]{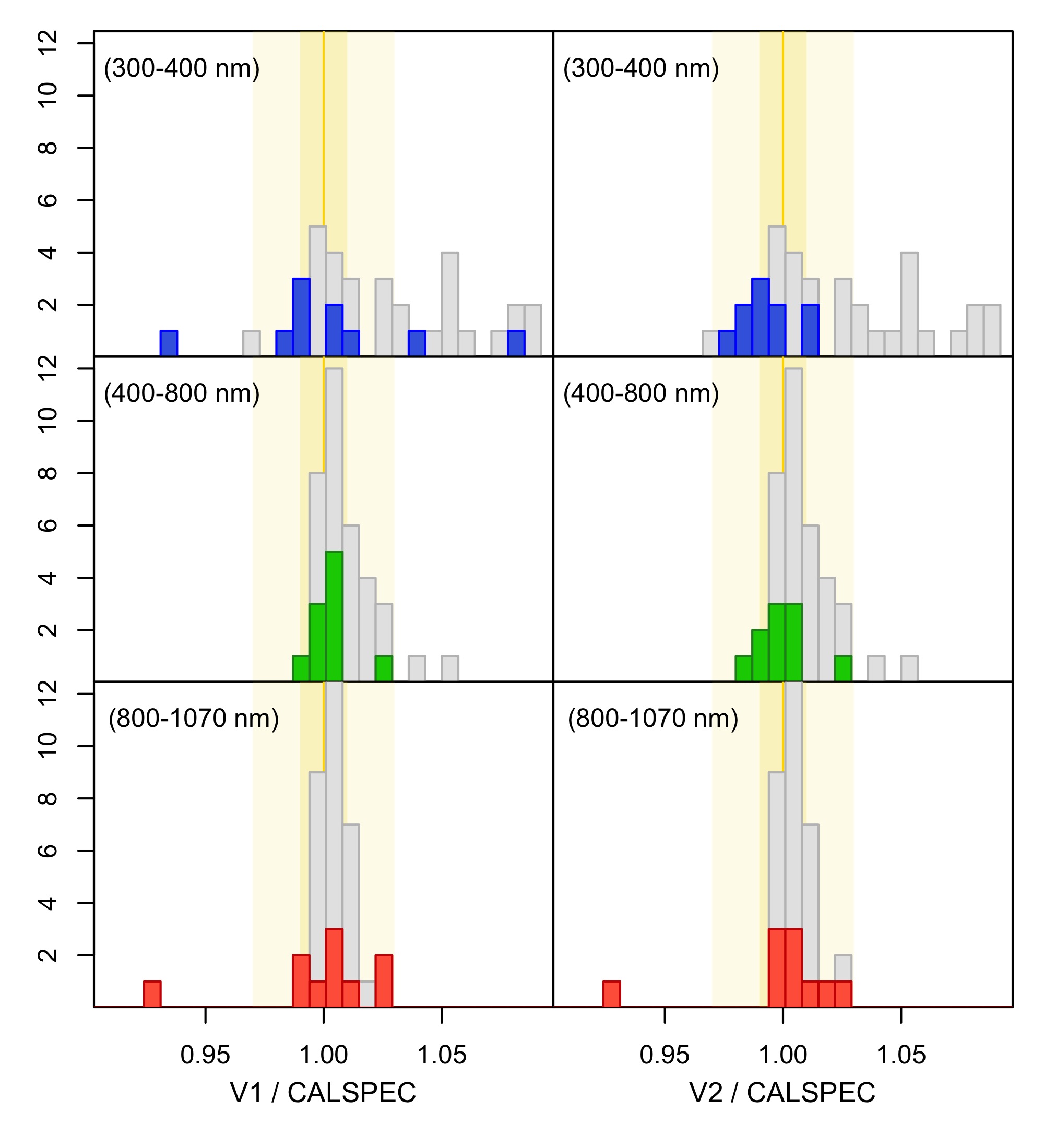} 
\vspace{-0.3cm}
\caption{Ratio of the SPSS V1 (left panels) and V2 (right panels) flux tables with the CALSPEC ones \citep{bohlin19}, for the 11 stars in common. The top panels show the median ratio in the 300--400~nm range (in blue), the middle panels in the 400--800~nm range (in green), and the bottom ones in the 800-1070~nm range (in red), as labelled. The ratio of the most recent releases of the CALSPEC spectra of the same star, when available, are plotted in grey, to give an idea of the CALSPEC internal uncertainty over time. Perfect agreement is marked by an orange vertical line, and the $\pm$1\% and $\pm$3\% agreement ranges are highligted in shades of orange in all panels (see Table~\ref{tab:calspec}).} \label{fig:calspec}   
\end{figure}
\begin{table*}
\centering
\caption{Synthetic magnitudes in the Johnson-Kron-Cousins and the {\em Gaia} EDR3 systems, computed on the SPSS V2 flux tables (see Section~\ref{sec:comp} for details). Only the first five rows are shown here, the full table is available online (see Data Availability Section).}\label{tab:syn}
\begin{tabular}{|l|c|c|c|c|c|c|c|c|}
\hline
SPSS & U & B & V & R & I & G & G$_{\rm{BP}}$ & G$_{\rm{BP}}$ \\
 & (mag) & (mag) & (mag) & (mag) & (mag) & (mag) & (mag) & (mag) \\
\hline 
1 & 10.247$\pm$0.005 & 11.482$\pm$0.006 & 11.790$\pm$0.008 & 11.931$\pm$0.012 & 12.112$\pm$0.021 & 11.721$\pm$0.009 & 11.541$\pm$0.006 & 12.067$\pm$0.019 \\

2 & 11.667$\pm$0.005 & 12.804$\pm$0.010 & 13.041$\pm$0.013 & 13.175$\pm$0.016 & 13.338$\pm$0.031 & 13.000$\pm$0.014 & 12.844$\pm$0.010 & 13.302$\pm$0.026 \\

3 & 11.908$\pm$0.004 & 13.088$\pm$0.008 & 13.355$\pm$0.011 & 13.495$\pm$0.011 & 13.671$\pm$0.021 & 13.304$\pm$0.011 & 13.136$\pm$0.008 & 13.628$\pm$0.019 \\

5 & 10.689$\pm$0.011 & 11.403$\pm$0.008 & 11.373$\pm$0.006 & 11.459$\pm$0.012 & 11.527$\pm$0.018 & 11.405$\pm$0.010 & 11.345$\pm$0.008 & 11.528$\pm$0.018 \\

6 & 12.583$\pm$0.007 & 13.798$\pm$0.010 & 14.067$\pm$0.012 & 14.206$\pm$0.019 & 14.369$\pm$0.037 & 14.013$\pm$0.015 & 13.845$\pm$0.010 & 14.335$\pm$0.031 \\
\hline
\end{tabular}
\end{table*}

\begin{figure} 
\includegraphics[clip,width=\columnwidth]{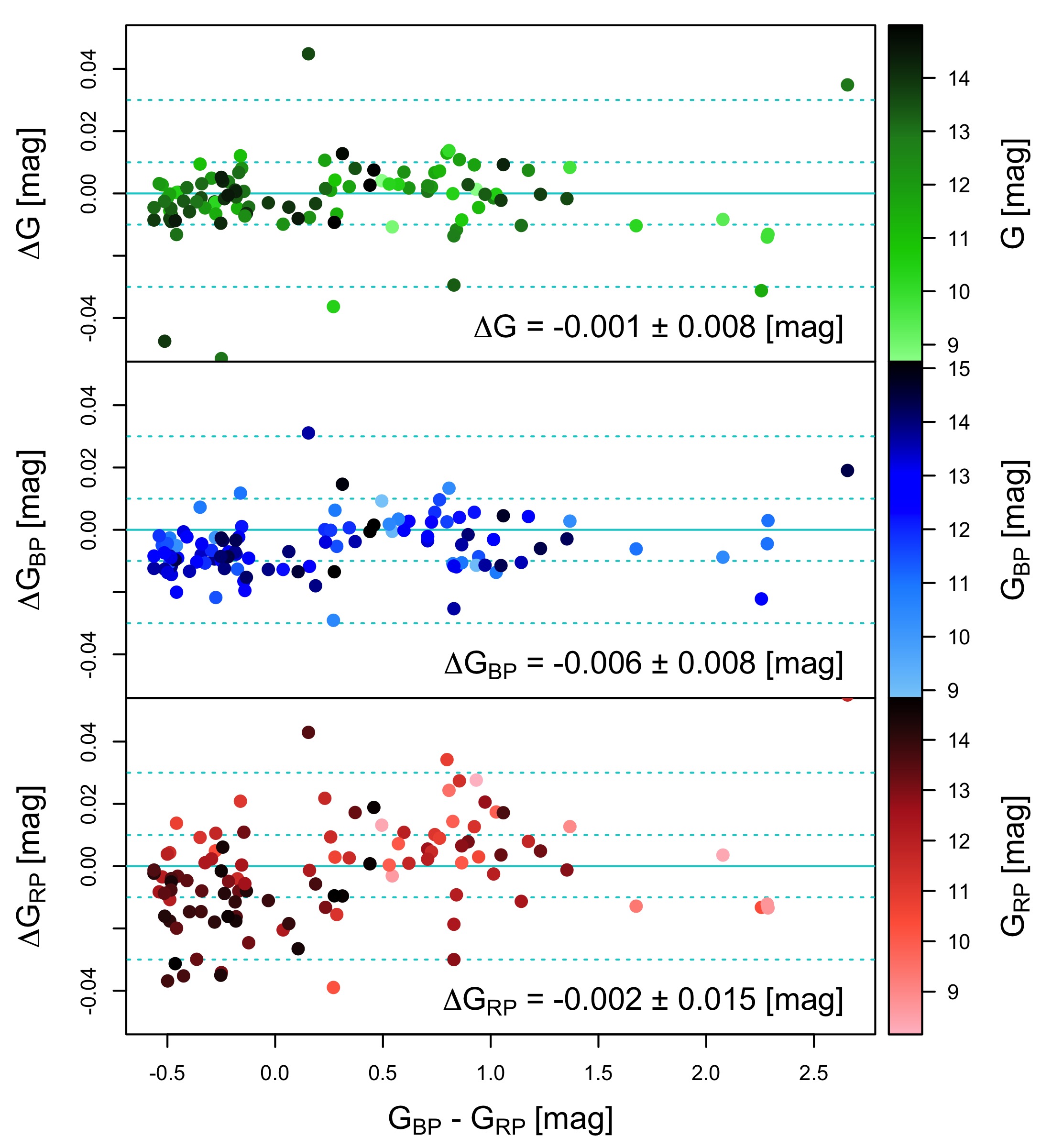} 
\vspace{-0.3cm}
\caption{Difference between the synthetic magnitudes computed on the SPSS V2 flux tables (Section~\ref{sec:comp}) and the {\em Gaia} EDR3 integrated magnitudes, as a function of the G$_{\rm{BP}}$--G$_{\rm{RP}}$ color. The SPSS are colored as a function of their {\em Gaia} EDR3 observed G magnitude, as indicated in the legend. The top panel shows the case of G, the middle one of G$_{\rm{BP}}$, and the bottom one of G$_{\rm{RP}}$. Perfect agreement is indicated by a solid cyan line in all panels, while the $\pm$1 and $\pm$3\% agreements are indicated by dotted lines.}
\label{fig:syngaia}   
\end{figure}

\subsection{SPSS literature comparisons}
\label{sec:comp}

\subsubsection{CALSPEC spectra}
\label{sec:calspec}

As mentioned in Section~\ref{sec:calib}, the SPSS flux reference system is tied to the CALSPEC system as it was in 2013. After that, the CALSPEC system was adjusted by making it 0.6\% fainter on average \citep{bohlin14}. More recently, the CALSPEC system was re-adjusted again by making it 0.87\% brighter \citep{bohlin20}. As a result, we expect our SPSS grid to agree, within uncertainties, with the latest CALSPEC release. We thus compared the SPSS V1 and V2 datasets with the latest version of the CALSPEC library, with the results reported in Table~\ref{tab:calspec} and in Figure~\ref{fig:calspec}. We note that in the bluest range (300--400~nm) the SPSS spectra are about 1\% fainter than the CALSPEC ones. We can also note that the SPSS V2 release shows a smaller scatter in the blue than the V1 release when compared with the same set of CALSPEC spectra (Table~\ref{tab:calspec} and Figure~\ref{fig:calspec}), thus the precision of the blue side of the SPSS spectra has improved between the two releases, likely because of the increased number of spectra used. In the central range (400--800~nm), where the SPSS spectra have the highest S/N ratio, the SPSS and CALSPEC fluxes agree to better than 0.5\%, as expected. We note that V1 had a slightly better agreement with CALSPEC than V2, especially at longer wavelength, albeit with a larger scatter. In preparation for the next SPSS release, we are carefully revising all the reduction and calibration steps to understand the exact cause for the change. This slight worsening of the median ratio above 800~nm might in fact be related to the $\simeq$1\% offset observed by \citet[][Figure~25]{riello20} for faint blue stars. We can finally note from Figure~\ref{fig:calspec} that the latest releases of the spectra in the CALSPEC library for the same star (plotted in grey in the background) differ from each other by amounts that are comparable to the scatter in the comparison between the SPSS and CALSPEC libraries. 

\begin{figure} 
\includegraphics[clip,width=\columnwidth]{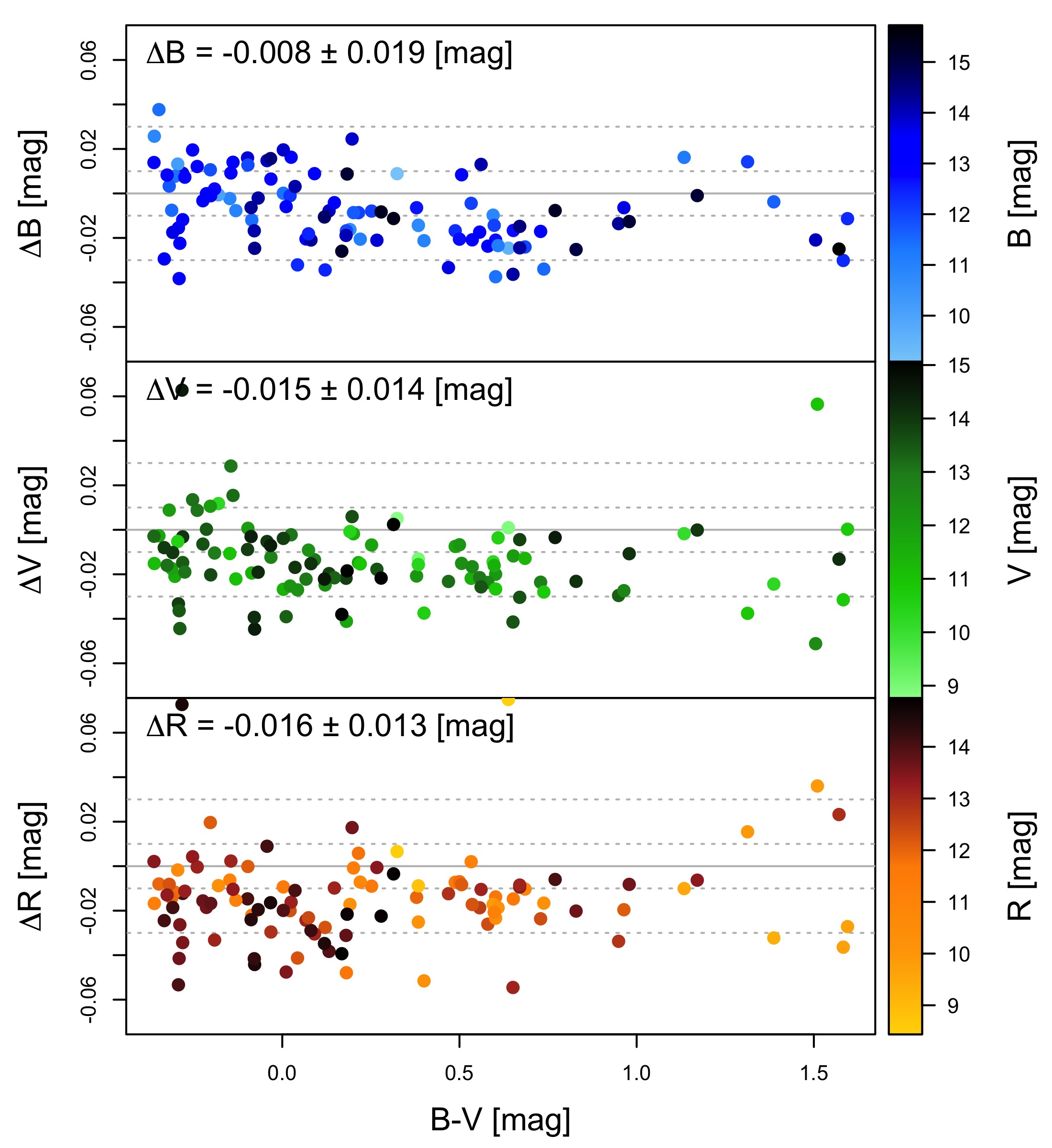} 
\vspace{-0.3cm}
\caption{As Figure~\ref{fig:syngaia}, but comparing with the Johnson-Kron-Cousins magnitudes of the SPSS by \citet{altavilla20}. The top panel refers to the B magnitude, the middle one to V, and the bottom one to R, as labelled.}
\label{fig:beps}   
\end{figure}

\subsubsection{Gaia EDR3 magnitudes}
\label{sec:gaia}

As a basic sanity check, we computed {\em Gaia} synthetic magnitudes on the SPSS V2 spectra and compared them with the corresponding {\em Gaia} EDR3 integrated magnitudes \citep{riello20}. We used the EDR3 passbands\footnote{\url{https://www.cosmos.esa.int/web/gaia/edr3-passbands}} and the same Vega spectrum used by \citet{riello20}, for consistency. The results of the comparison are shown in Figure~\ref{fig:syngaia}. A number of outliers is clearly present. These will be individually examined once the full set of spectra will be reduced, to identify and solve any obvious problem in the reduction chain. The median offsets and median absolute deviations, computed in the sense of SPSS synthetic magnitudes minus {\em Gaia} EDR3 ones, are: $\Delta$G=--0.001$\pm$0.008, $\Delta$G$_{\rm{BP}}$=--0.006$\pm$0.008, and  $\Delta$G$_{\rm{RP}}$=--0.002$\pm$0.015~mag. An excellent agreement is found, with a slight negative bias of the order of 0.1--0.6\%, i.e., negligible compared to the uncertainties. A larger spread is observed in the comparison with G$_{\rm{RP}}$, with a possible trend with magnitude among blue stars. This might in part be related to the above mentioned 1\% differences found between SPSS V1 and V2 redward of 800~nm, in the sense that some extra uncertainty appears to be present in the SPSS flux tables of faint blue stars (darker symbols in the bottom panel of Figure~\ref{fig:syngaia}). As mentioned, this effect was also noted by \citet[][Figure~25]{riello20}. If {\em Gaia} magnitudes were significantly affected by this minor SPSS defect, we would observe a better agreement in the bottom panel of Figure~\ref{fig:syngaia}. We also note some jumps at about G$_{\rm{BP}}$--G$_{\rm{RP}}$ that are not present in the comparisons with CALSPEC or the ground-based SPSS photometry.

\begin{figure} 
\includegraphics[clip,width=\columnwidth]{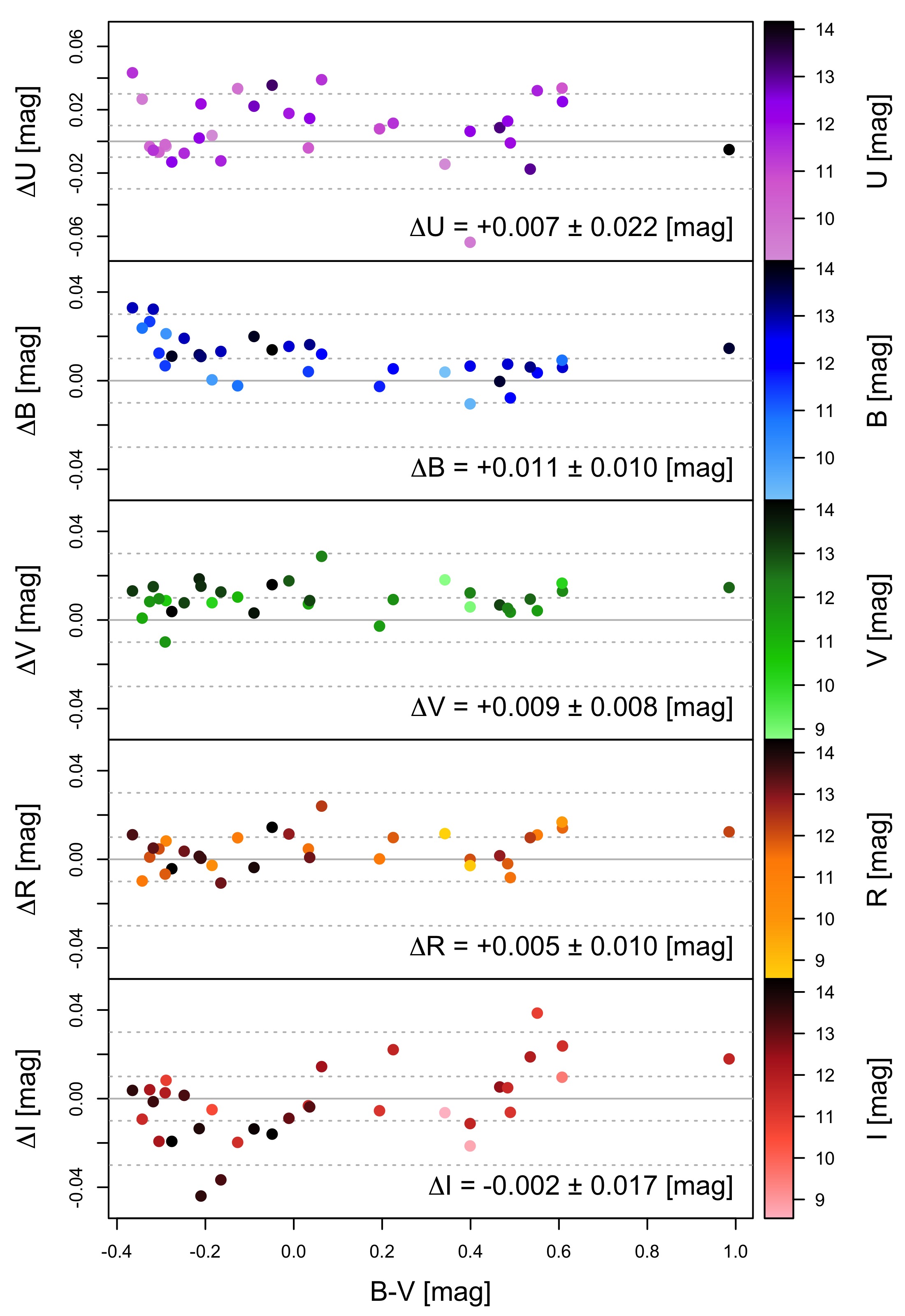} 
\vspace{-0.3cm}
\caption{As Figures~\ref{fig:syngaia} and \ref{fig:beps}, but comparing with the Johnson-Kron-Cousins magnitudes from a series of papers by Landolt and collaborators (see Section~\ref{sec:comp}). Each panel refers to a different passband, as labelled.}
\label{fig:landolt}   
\end{figure}

\subsubsection{Johnson-Kron-Cousins magnitudes}
\label{sec:landolt}

We also computed the synthetic Johnson-Kron-Cousins UBVRI magnitudes on the SPSS V2 set and compared them with our own absolutely calibrated magnitudes \citep{altavilla20} and, for 36 stars, with the original measurements in a series of papers by Landolt and collaborators \citep{landolt83,landolt92b,landolt07,landolt09,bohlin15}. To compute the Johnson-Kron-Cousins magnitudes, we used the passbands by \citet{bessell12}. The choice of the Vega spectrum is crucial, as variations of 1--2\% can be easily obtained in the results when using different spectra. We thus employed the same Vega spectrum used for the {\em Gaia} comparison above (Section~\ref{sec:gaia}), i.e., the Kurucz spectrum of an A0V star, normalized at 550~nm to a reference flux of 3.660~10$^{-11}$~W~m$^{-2}$~nm$^{-1}$, according to \citet{straizys92}. If we recompute the magnitude on the Vega spectrum {\tt alpha\_lyr\_stis\_005} used by \cite{bessell12}, we obtain U=0.034,  B=0.025, V=0.027, R=0.028, and I=0.028~mag, in good agreement (better than 1\%) with their results. This ensures that our synthetic magnitudes are on the standard scale, within uncertainties.

The comparison of the resulting synthetic magnitudes, obtained on the SPSS V2 spectra, with the magnitudes by \citet{altavilla20} is shown in Figure~\ref{fig:beps}. As can be observed, the scatter is larger than in the case of Figure~\ref{fig:syngaia}, confirming the excellent quality and homogeneity of the {\em Gaia} EDR3 photometry. The R magnitude comparison hints at a  segregation of faint blue stars (darker points in the bottom panel of Figure~\ref{fig:beps}, as noted in the previous section), which appear to have a slightly lower average difference. The SPSS V2 synthetic magnitudes appear about 1--2\% brighter than the \citet{altavilla20} ones, which in turn appeared to be about 1.5\% fainter than the Landolt measurements (see their Figure~8). This is confirmed when comparing our SPSS V2 synthetic magnitudes with the ones from the Landolt collection in Figure~\ref{fig:landolt}: the SPSS V2 magnitudes agree with the ones of the Landolt collection to better than 1\%. Given the involved uncertainties in the magnitude measurements \citep[see][]{clem13} and in the synthetic magnitudes computation, the agreement can be considered quite satisfactory. In this case, the R magnitude comparison appears excellent  for the faint blue stars, while some problems appear in the I band comparison, where the scatter is large and again the faint blue stars appear to behave slightly differently from the other SPSS. This suggest that the problems in the SPSS V2 spectra of faint blue stars, if any, should start after $\simeq$700--800~nm. We also note the large scatter in the U band comparison: the U band is notoriously difficult to standardize \citep{bessell05,stetson00,stetson19}. This sums up with the lower S/N ratio typically reached in the SPSS observed spectra below $\simeq$400~nm. As shown in the comparison between SPSS V1 and V2 (Section~\ref{sec:calspec}), increasing the number of spectra can significantly improve the scatter in the blue. Further improvement in this sense is expected in the next SPSS release. 

\subsection{Expected improvements in the next SPSS release}
\label{sec:v3}

As already mentioned, the current SPSS release is still preliminary, although it already meets the original requirements, showing an excellent agreement with both the CALSPEC and Landolt reference systems. The next release, SPSS V3, that will be used to calibrate {\em Gaia} DR4, is expected to further improve based on the following considerations:

\begin{itemize}
\item{the entire set of about 6500 spectra will be analyzed, increasing the S/N ratio and smoothing out further pixel-to-pixel defects and residuals of the various advanced spectra reductions (section~\ref{sec:reds});}
\item{the entire set of surviving SPSS candidates, which at the moment are about 200, will be included; the list of candidates will be further cleaned from variable and multiple stars (Section~\ref{sec:sea});}
\item{the source of the $\simeq$1\% offset observed on the red part of the spectrum ($\gtrsim$700--800~nm) for faint blue SPSS (Section~\ref{sec:comp}) will be investigated and repaired if possible;}
\item{the calibration process used in V1 and V2 was tied to the 2013 version of the CALSPEC grid (Section~\ref{sec:calib} and Table~\ref{tab:calibs}), while in V3 we plan to calibrate onto the latest CALSPEC spectra of the three pillars only, following the Pillar-Primary-Secondary calibration scheme \citep[see also Section~\ref{sec:calib}]{pancino12};}
\item{the extension of the observed flux tables with template spectra to replace the noisy borders will be updated with new template spectra and more refined algorithms, to minimize discrepancies at the junctions (ideally within 1\% or less).}
\end{itemize}

\begin{table*}
\caption{List of stars included in the PVL along with basic information: the PVL ID and name, the {\em Gaia} EDR3 ID and coordinates (rounded), the  literature B and V magnitudes and the spectral types from the original literature sources of the spectra. Only the first five lines are shown here, the table is available in its entirety online (see Data Availability Section).}\label{tab:pvllist}
\begin{tabular}{|l|l|l|r|r|r|r|r|r|l|}
\hline
PVL ID  &  Name   & {\em Gaia} EDR3 ID & RA & Dec & B & V & Sp. Type & Source & Notes \\
& & & (deg) & (deg) & (mag) & (mag) & & & \\
\hline                                401	& BD\,--11\,3759 & 6324325225803432320	& 218.56848 & --12.51701 & 12.9 & 11.3 & M3.5V & CALSPEC & Flare star \\
402	& Prox Cen &	5853498713190525696	& 217.39347 & --62.67618 & 12.9 & 11.1 & M5.5V & \citet{ribas17} & Flare star; dust rings \\
403	& V\,B8	& 4339417394313320192 & 253.89336 & --8.39840 & 18.7 & 	16.92 & M7V & CALSPEC & Flare star; variable \\
404	& 2M0036+18 & 2794735086363871360 & 9.06713 & +18.35286 & --- & --- & L3.5 & CALSPEC & Brown dwarf \\
405	& 2M0559-14 & 2997171394834174976 & 89.82995 & --14.08034 & --- & --- & T4.5 & CALSPEC & Brown dwarf \\
\hline
\end{tabular}
\end{table*}

All these improvements are not expected to significantly change the overall accuracy (i.e., the zeropoint) of the SPSS grid, but are expected to improve its precision (i.e., the homogeneity). In fact, the accuracy of the SPSS V2 preliminary release already meets the initial requirements set within the {\em Gaia} DPAC and is of the order of 1\% or better, which is considered the current technological limit of flux calibrations \citep{clem13,bohlin14,altavilla20}. The precision, on the other hand, can be improved both {\em within SPSS} (flux tables precision) and {\em among SPSS} (grid homogeneity). Within SPSS, by increasing the S/N ratio and by reconciling the blue and red edges of each flux table with the high-quality central parts.  Among SPSS, by calibrating more homogeneously and self-consistently each SPSS on the three CALSPEC pillars only, instead of the larger collection of calibrators listed in Table~\ref{tab:calibs}, and by investigating the cause of the $\simeq$1\% offset in the red part of the flux tables ($\gtrsim$700--800~nm) of the blue and faint stars.

\begin{figure} 
\includegraphics[clip,width=\columnwidth]{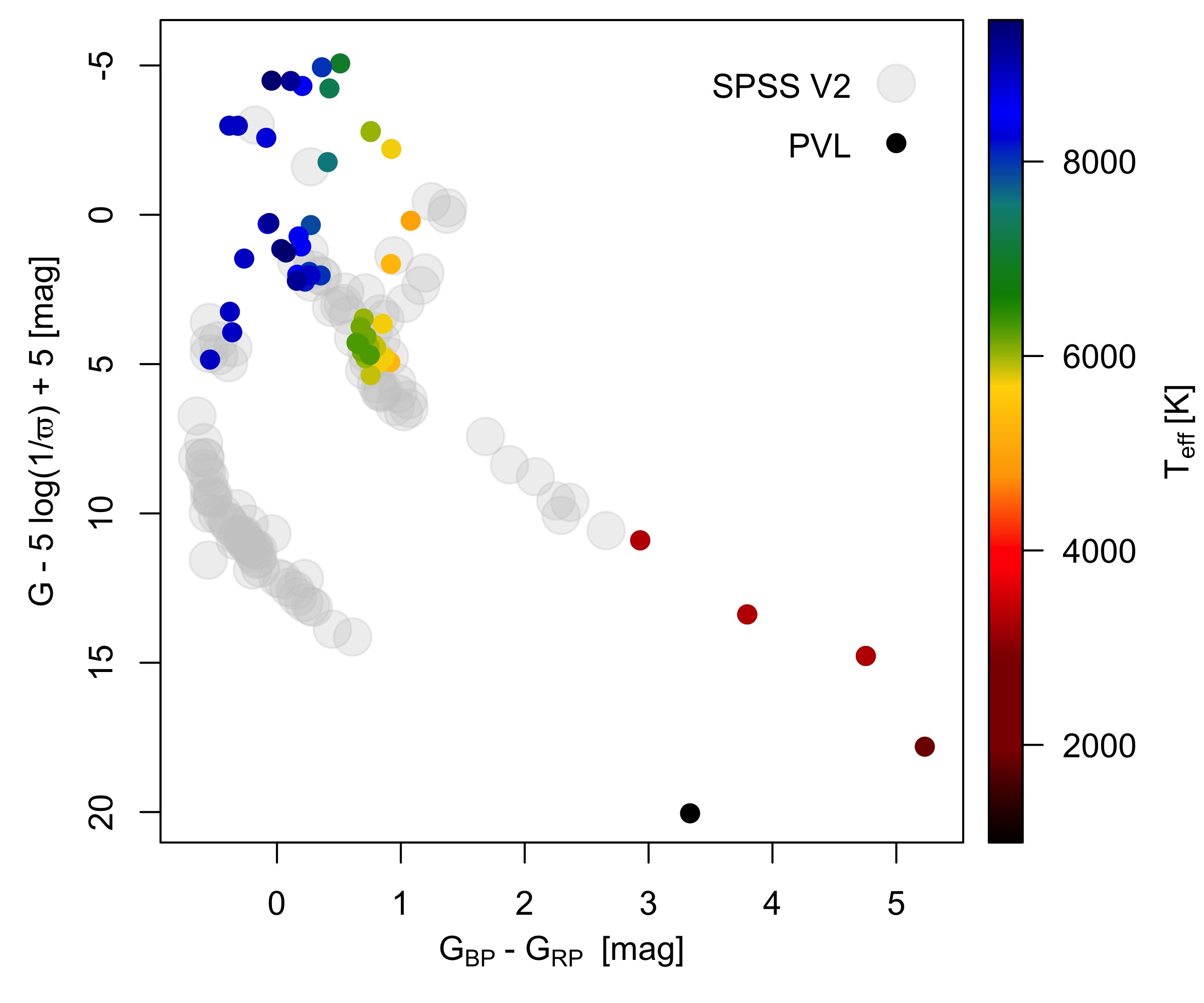} 
\vspace{-0.3cm}
\caption{The {\em Gaia} colour -magnitude diagram of the PVL. The SPSS V2 stars are plotted in the background as grey circles for comparison. The PVL stars are colored as a function of their T$_{\rm{eff}}$ (the color scale saturates at T$_{\rm{eff}}>9000$~K, for readability).}
\label{fig:pvlcmd}   
\end{figure}

\section{The passband validation library (PVL)}
\label{sec:pvl}

After the publication of the {\em Gaia} DR2 photometry \citep{evans18}, a variety of passbands and tests on the quality of the integrated G, G$_{\rm{BP}}$, and G$_{\rm{RP}}$  magnitudes appeared in the literature \citep{evans18,casagrande18,weiler18,maiz18}. In particular, given the complexity of the multiparametric model for the (internal and external) flux calibration of {\em Gaia} data, it was suggested that the lack of variety in stellar spectra in general -- and in the SPSS set in particular -- might be the cause of some small systematics observed in the {\em Gaia} DR2 photometry \citep{maiz18} and in the differences among some of the published passbands. A library of well calibrated spectra covering also extreme spectral types, that do not match our criteria as flux calibrators (especially flux constancy, see next sections), was assembled by \citet{maiz18}, using mostly HST spectra, including a populous sample of O and B stars and three very red stars. Unfortunately, that library of spectra is not publicly available. We thus assembled a separate library, the PVL, using the original data sources cited by \citet{maiz18}, whenever possible, as described in the following. The error estimates on the flux were computed by summing in quadrature, when known, the formal flux uncertainties, the flux variability, and additional systematic uncertainties. We made the effort, whenever appropriate, to place the PVL on the same flux reference system of the SPSS (see also Section~\ref{sec:comp}) so that the two sets can be used together for a variety of validation tasks.

\subsection{PVL stars from CALSPEC}
\label{sec:pvlcalspec}

We selected a list of 39 stars from the 2019 CALSPEC version\footnote{\url{ftp://ftp.stsci.edu/cdbs/current_calspec}} \citep{bohlin19} by
excluding: {\em (i)} the 11 stars already included in the SPSS V2 set\footnote{In reality, star
1757132 was added to the CALSPEC set in 2015, and we did not realize that it
corresponds to SPSS 309 (2MASSJ17571324+6703409), that was added to the SPSS
candidates in 2012 as one of the stars close to the North Ecliptic Pole for {\em Gaia}
GDR1 testing. We decided to leave the star in PVL as a comparison object with the SPSS
set.}; {\em (ii)} all the WDs, which are already well represented in the SPSS V2 set;
{\em (iii)} stars with incomplete STIS coverage; {\em (iv)} variable stars; {\em (v)} stars with
problems related to SNR, resolution, or spectral stitching; {\em (vi)} stars that are
too bright to be observed accurately by {\em Gaia} (e.g., Sirius and $\alpha$ Lyr).
Four CALSPEC very red stars are treated separately in Section~\ref{sec:pvlred}. The
CALSPEC set adds 15 A-type stars, 3 B-type stars, 1 O-type star, 6 F-type
stars, 11 G-type stars, and 3 subdwarfs to the SPSS V2 coverage. For each of the 39 selected CALSPEC stars, we selected the latest and most
reliable CALSPEC
spectrum. We brightened
the flux of each spectrum by 0.6\% to bring it to the pre-\citet{bohlin14} CALSPEC system, on which the SPSS V2 is calibrated\footnote{All these selections were made in 2019, i.e., before the latest revision of the CALSPEC grid \citep{bohlin19,bohlin20}, thus the spectra are on average calibrated on a system that is 0.6\% brighter than the one of the SPSS, based on the 2013 CALSPEC version.}, and we
combined the statistical and systematic errors in quadrature to provide one
single error estimate. The spectra were not extended with models because they cover the entire range
300-1200~nm, except for star HD\,60753, for which the spectrum stops at about
1022~nm, but this is a hot star with very little flux in the red. 

\subsection{PVL hot and bright stars}
\label{sec:pvlhot}

As mentioned, hot and bright stars were only included in small numbers in the SPSS list due to the high fraction of variable and binary stars \citep{demarco17,eyer19}. Therefore, they were important to include in the PVL.
The \citet{maiz18} compilation contained hot and bright stars from two separate
sources, that they labelled ``Massa" and ``Hot" star sets. The former set is not
publicly available, while the latter was published by \citet{khan18} as a hot
star extension of the NGSL \citep[New Generation Spectral Library, ][]{ngsl}, containing 41 stars in the post-AGB
phase, HB stars, and O-type stars, 17 of which were considered by \citet{maiz18}.
We thus examined the star list in \citet{khan18} and decided to: {\em (i)} avoid all
the planetary nebulae and proto-PN of the set, including one proto-PN that was
in \citet{maiz18} and is variable; {\em (ii)} discard all WDs, that are well represented
in the SPSS V2 set; {\em (iii)} remove all variable stars, including 3 stars that were
part of the \citet{maiz18} list. The final list contains 17 stars, of which 14 are
in common with the \citet{maiz18} set.
The flux calibration of these spectra is not well described in the original
paper \citep{khan18}, but reference is made to NGSL \citep{ngsl}, which is
calibrated with an accuracy of about 3\%. We thus added an error component of
3\% of the flux to compute the final error budget, and we did not apply the
(comparatively small) 0.6\% flux correction, because it is not clear on which
version of the CALSPEC system these stars were calibrated. STIS
spectra reach 1025~nm approximately, but these hot stars have very little flux
at the red edge and therefore the spectra were not extended with models.

\subsection{PVL very red stars}
\label{sec:pvlred}

Very red stars are known to be variable because of chromospheric activity,
rotation, star spots, magnetic activity, and flares. Brown dwarfs are known to be variable because of clouds complexes in their atmospheres, modulated by rotation and seasonal changes \citep{artigau18}. Therefore, when assembling our sample of very red stars, we took into account the known variability found in the literature, and included it in the error computation, as detailed below. 
We selected from CALSPEC four very red stars: GJ\,555, VB\,8, and two brown dwarfs (2M0036+18 and 2M0559-14). The two brown dwarfs have no observations blueward of 500~nm, because in that range they emit very little flux. Nevertheless, they were detected by {\em Gaia}. We extended their spectra with the appropriate models from the \citet{allard12} grid. All the spectra were treated like the CALSPEC spectra in Section~\ref{sec:pvlcalspec}: we selected the latest spectrum and brigthened to the level of the pre-\citet{bohlin14} CALSPEC reference system, on which the SPSS V2 is calibrated. We used the typical flux variations of the respective spectral types \citep{artigau18} as the flux error component caused by possible brown dwarf variability, amounting to 2\% for 2M0036+18 and to 1\% for 2M0559-14. For VB\,8 we used the estimated variability by \citet{martin96} to add a 2\% component to the flux uncertainty. For GJ\,555, we used the variability estimated by \citet{hosey15} of 1.4\%. 

We further added Proxima Centauri, that was part of the \citet{maiz18} grid. We used the spectrum by \citet{ribas17}, although the uncertainties in this case are high: {\em (i)} the various spectra used by \citet{ribas17} agreed with each other within 2--3\%; {\em (ii)} the sources of flux variability examined by \citet{ribas17} amount to about 5\%; and {\em (iii)} the authors used STIS spectra taken with a narrow slit (0.2"). We thus conservatively estimated the uncertainty of this spectrum to be about 5\%. This star should be used with caution also because of a suspect disk of warm polarized dust, but mostly because of flares \citep{macgregor18}, and in fact it displays a very large scatter in Figures 6 to 9 by \citet{maiz18}. We also note that Proxima Centauri was not included in the CALSPEC grid, presumably for the above reasons.

\begin{figure} 
\includegraphics[clip,width=\columnwidth]{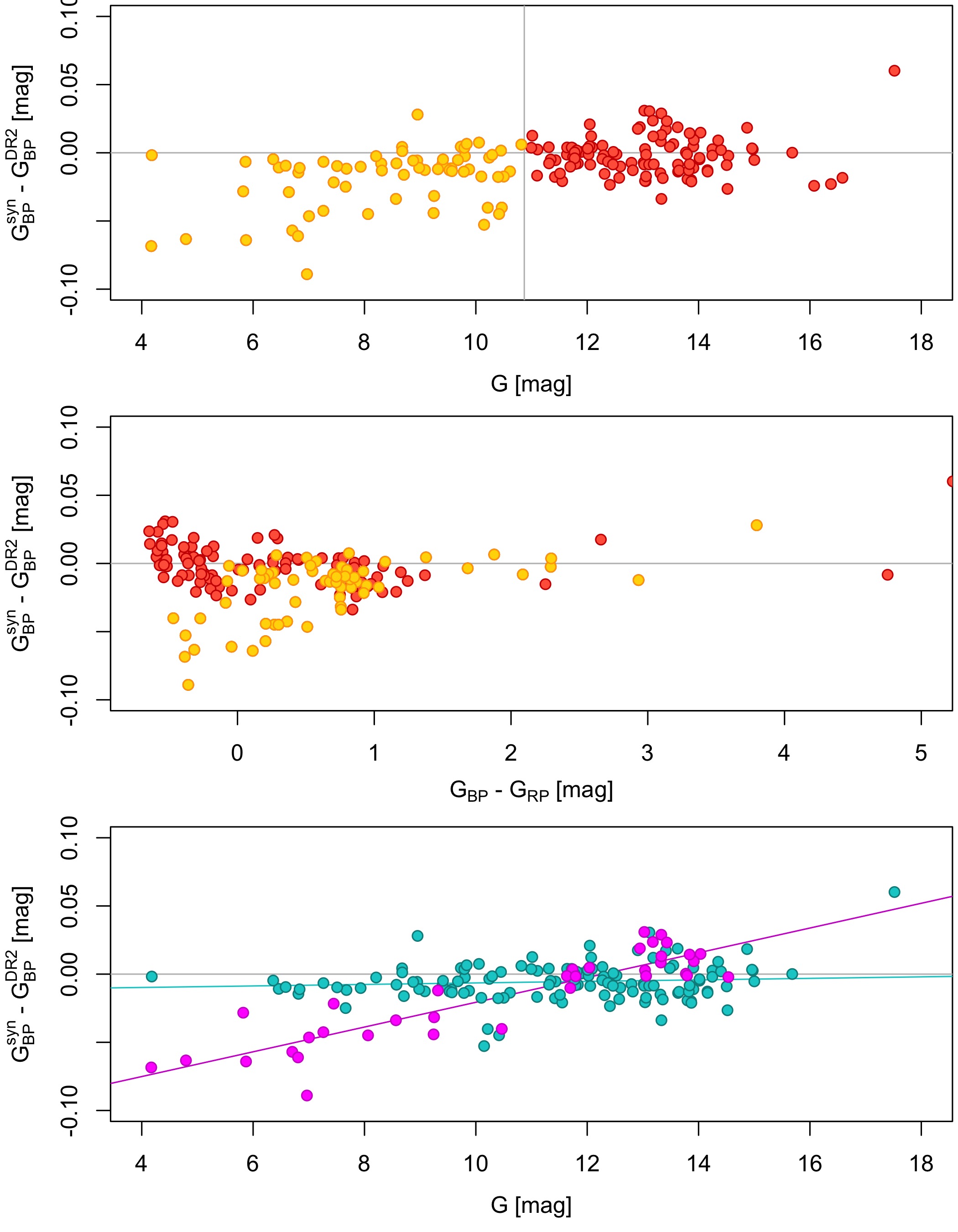} 
\vspace{-0.3cm}
\caption{Example of how the SPSS + PVL library was used to investigate the problems of bright blue stars in {\em Gaia} DR2, that are now much less prominent in EDR3. {\em Top panel.} The difference between the G$_{\rm{BP}}$ synthetic magnitude and the DR2 one for the global SPSS + PVL sample. The cut at G=10.87 proposed by \citet{maiz18} is marked by a vertical line: fainter stars are colored in red and brighter ones in gold. {\em Middle panel.} Same as the top panel, but as a function of G$_{\rm{BP}}$--G$_{\rm{RP}}$ instead of G. Fainter stars (in red) have on average consistent G$_{\rm{BP}}$, while bright blue stars (in gold) show a fan-shaped discrepancy at blue colors. {\em Bottom panel.} Same as the top panel, but now stars are colored according to T$_{\rm{eff}}$ rather than G magnitude. Linear fits to the samples with T$_{\rm{eff}}$ lower (in cyan) and higher (in magenta) than 30\,000~K are also plotted. In all panels, perfect agreement is marked by a horizontal grey line. }
\label{fig:pvlvalid}   
\end{figure}

\subsection{PVL content and results}
\label{sec:pvluse}

The PVL library contains 61 stars. Its content is illustrated in Figure~\ref{fig:pvlcmd} and the list of stars is presented in Table~\ref{tab:pvllist}. The flux tables are available online (see Data Availability Section) and the files are in the same format as the SPSS ones. As can be noted, the PVL contains many more hot and bright O and B stars, three more hot subdwarfs, a few FGK stars, and five extremely red stars, covering a range that is not included in the SPSS set (see also Figure~\ref{fig:cmd}). As mentioned, the two sets were calibrated as much as possible on the same flux reference system, so that they could safely be used together in a variety of validation procedures during the preparation of {\em Gaia} EDR3 \citep[see also][]{riello20}. 

As an example, the PVL was instrumental in investigating the apparent discrepant behaviour of blue stars found in DR2 by \citet{weiler18} and \citet{maiz18}. These authors noted that the blue stars fainter than G$\simeq$10.87~mag behaved differently from those brighter than that limit, and postulated the need for two different sensitivity curves to reconcile the magnitudes of bright and faint blue stars. We used the combined PVL and SPSS V2 global sample to explore the problem from a different point of view. We computed synthetic magnitudes using the same Vega spectrum used in Section~\ref{sec:gaia} and the REV passbands by \citet{evans18}. As shown in Figure~\ref{fig:pvlvalid}, while faint blue stars do have consistent synthetic and observed G$_{\rm{BP}}$ magnitudes, bright blue stars do belong to both the branches of an apparent bifurcation for colors bluer than G$_{\rm{BP}}$--G$_{\rm{RP}}$ of about 1~mag. In practice, there are bright blue stars that are well-behaved and other bright blue stars that do not behave well (middle panel of Figure~\ref{fig:pvlvalid})\footnote{This is slightly different from the \citet{maiz18} conclusion that all the bright blue stars behave differently from faint blue ones.}. This is true regardless of the exact cut in magnitude: we used values ranging from G=10.5 to 12.5~mag obtaining qualitatively the same result (top panel of Figure~\ref{fig:pvlvalid}). However, we realized that hot stars with T$_{\rm{eff}}>30\,000$~K lie on the discrepant branch, while hot stars cooler than 30\,000~K lie on the well behaved branch. Moreover, faint WDs hotter than 30\,000~K appear to lie on the opposite side of the zero line (bottom panel of Figure~\ref{fig:pvlvalid}), suggesting that these extreme spectral types, that are not so numerous, were not well represented in the internal calibration model used for DR2. As can be noted in Figure 5.3 in Section 5.1.3 of the {\em Gaia} online DR2 documentation\footnote{\url{https://gea.esac.esa.int/archive/documentation/GDR2/}}, the shape of BP spectra changes above $\simeq$30\,000~K in the blue side of the spectra. As a result of this and many other investigations, the systematic effects in {\em Gaia} ERD3 are now greatly reduced compared to the ones in DR2. Further improvement is expected in future releases, as the amount of data increases as well as our knowledge and understanding of the data properties.  

\section{Discussion and conclusions}
\label{sec:concl}

We presented the flux tables for the first two preliminary releases of the {\em Gaia} SPSS: the V1, used to calibrate {\em Gaia} DR2 and the V2, used to calibrate {\em Gaia} (E)DR3. We complemented the SPSS grid with the PVL, a library of stars that do not pass all of our criteria to be flux standards, but which possess well calibrated spectra in the literature. Most importantly, the PVL contains very red and bright blue stars that are normally not used in great quantities as flux calibrators. The SPSS + PVL combined set was used as a validation set during the processing of {\em Gaia} (E)DR3 photometry. 

The SPSS preliminary releases already meet the required accuracy for the {\em Gaia} project in the flux reference system: they agree to better than 1\% with the CALSPEC grid for 11 stars in common (Section~\ref{sec:calspec}) and with a collection of 37 stars from a series of papers by Landolt and collaborators (Section~\ref{sec:landolt}). The current SPSS releases do not contain all the observed SPSS spectra and all the surviving SPSS candidates, thus improvements are expected in the upcoming SPSS release, that will be used to calibrate {\em Gaia} DR4 (Section~\ref{sec:v3}). Concerning the internal homogeneity of the SPSS grid, we observe a scatter of the order of 1\% in various literature comparisons, but the current SPSS V2 version shows a slightly discrepant ($\simeq$1\%) behaviour for faint blue stars in the red part of the SPSS flux tables ($\gtrsim$700~nm) that will be investigated in future releases.

The current SPSS release is tied to the CALSPEC grid, as it was in 2013. The CALSPEC reference system oscillates by 0.5--2.0\% from release to release \citep[][see also Section~\ref{sec:calib}]{bohlin14,bohlin20}, but the latest three CALSPEC releases appear to converge on sub-1\% readjustments of the overall flux zero-point in the optical and near-IR range. One important limitation of the CALSPEC grid, and indeed of all the available sets of flux standards, is that very few direct flux measurements exist \citep{deustua13}. Thus the CALSPEC grid is tied to the best flux models available in the literature for the three WD pillars \citep{bohlin95}. WDs are relatively simple, their models are well understood and reliable, and the stars are rarely variable if one avoids the ZZ\,Ceti, V777~Her, and GW~Vir instability strips \citep{rauch14,hermes17}. However, an accurate and updated direct measurement of stellar fluxes is still lacking and very difficult to attain in practice. In other words, the vast majority of the spectro-photometric flux standards available in the literature (including the Landolt and CALSPEC systems) are in practice made of {\em secondary standards}, only indirectly based on actual standardized measurements. Few {\em primary standards} exist for the UV, optical, and near-IR ranges, such as Vega, \citep{selby83,blackwell83,megessier95} Sirius \citep[in the IR, ][]{price04}, and the Sun \citep{sun}.

Several astrophysical investigations, including the dark energy surveys based on observations of Type~Ia supernovae, do require an accurate (i.e., {\em absolute}) flux calibration to 0.5\% or better \citep{albrecht06}. Therefore, updated and numerous direct measurements of flux standards are a pressing need, to go as much as possible beyond the 1\% limit currently experienced by the best flux standard sets available. Vega, the best measured primary standard so far, displays $\simeq$4\% flux variations and an IR excess caused by a debris disk \citep{megessier95}. Therefore, Sirius has been proposed as a more reliable primary standard, but it can count on fewer standard flux measurements so far \citep{price04}. The "NIST stars"\footnote{\url{https://www.nist.gov/programs-projects/nist-stars}} is an ongoing project to measure the irradiance of a few reliable stars, known to vary less than 1\% in flux, in a dedicated and carefully designed experiment \citep{smith09}. To complement NIST stars, ACCESS \citep{kaiser17} is a project consisting in series of sub-orbital rocket and ground-based experiments, that will target a few stars, including Sirius and Vega, to improve the knowledge of their flux beyond the 1\% limit. Another US-funded experiment, ALTAIR \citep{albert15} will launch balloons with artificial sources to calibrate a few primary standards. The project is presently in a testing phase, evaluating a series of commercial photodiodes. 

In any case, for the moment, the main reference systems for the optical range remain CALSPEC for spectroscopy and Landolt for photometry. Our selection of SPSS -- optimized for the flux calibration needs of {\em Gaia} and homogeneously calibrated -- compares well to both systems, to the 1\% level or better. When new primary standards will be established and their calibration will be transferred to stars in the magnitude range of the SPSS, we will be able to refine our calibration, because the SPSS grid contains some of the most widely used spectro-photometric stars, such as the three CALSPEC pillars.

\section*{Data availability}

The SPSS and PVL flux tables are available though the dedicated SPSS@SSDC database and archive interface (\url{http://gaiaextra.ssdc.asi.it:8900}), together with all the relevant metadata and information. All the paper tables are available online as supplementary material to this paper on MNRAS. The data can also be obtained through the Vizier catalogue server (\url{http://vizier.u-strasbg.fr/viz-bin/VizieR}) at the CDS (\url{https://cds.u-strasbg.fr/}). 

\section*{Acknowledgements}

We would like to thank the following colleagues, who in various ways collaborated to the {\em Gaia} SPSS project, including past SPSS team members: V.~Braga, L.~Balaguer N\'{u}\~{n}ez, C.~Cacciari, A.~Castro, L.~Di Fabrizio, L.~Federici, F.~Figueras, F.~Fusi Pecci, M.~Gebran, C.~Lardo, E.~Masana, M.~Mongui\'o, P.~Montegriffo, M.A.~Pe\~{n}a-Guerrero, A.~P\'{e}rez-Villegas, M.~Riello, E.~Rossetti, W.J.~Schuster, S.~Trager, G.~Valentini, F.~Vilardell. We warmly thank the technical staff of the Calar Alto, Loiano, La Silla NTT, and Roque
de Los Muchachos TNG observatories. We would like to warmly thank M. Bessell, who has refereed several papers of the SPSS series, providing not only deep insight and useful comments, but also much needed encouragement.
The results presented here are based on observations made with: the ESO New Technology Telescope (NTT) in La Silla, Chile, (programmes: 182.D-0287(A); 182.D-0287(C); 086.D-0176; 0.91.D-0276; 0.93.D-0197; 094.D-0258); the Italian Telescopio Nazionale Galileo (TNG) operated on the island of La Palma by the Fundaci\'on Galileo Galilei of INAF (Istituto Nazionale di Astrofisica) at the Spanish ‘Observatorio del Roque de los Muchachos’ of the Instituto de Astrofisica de Canarias (programmes: AO/17\_TAC3; AOT19\_TAC13; AOT20\_TAC41; AOT21\_TAC1); the 2.2 m telescope at the Centro Astron\'omico Hispano-Alem\'an (CAHA) at Calar Alto, operated jointly by Junta de Andaluc\'\i a and Consejo Superior de Investigaciones Cient\'\i ficas (IAA-CSIC, programmes: F07-2.2-033; H07-2.2-024; F08-2.2-043; H08-2.2-041; F10-2.2-027; H10-2.2-0.42; F11-2.2-033; F12-2.2-034); the Cassini Telescope at Loiano, operated by the INAF, Astronomical Observatory of Bologna (7 observing runs between 2006 and 2015).

This work uses data from the European Space Agency (ESA) space mission {\em Gaia}. {\em Gaia} data are being processed by the Gaia Data Processing and Analysis Consortium (DPAC). Funding for the DPAC is provided by  national institutions, in particular the institutions participating in the {\em Gaia} Multi-Lateral Agreement(MLA). This research has made use of the GaiaPortal catalogues access tool, Agenzia Spaziale Italiana (ASI) - Space Science Data Center (SSDC), Rome, Italy (\url{http://gaiaportal.ssdc.asi.it}).
In particular, we would like to acknowledge the financial support of the Istituto Nazionale di Astrofisica (INAF) and specifically of the Arcetri, Roma, and Bologna Observatories; of ASI (Agenzia Spaziale Italiana) under the contract to INAF: ASI 2014-049-R.0 dedicated to SSDC, and under the contracts to INAF: ARS/96/77,ARS/98/92, ARS/99/81, I/R/32/00, I/R/117/01, COFIS-OF06-01,ASI I/016/07/0, ASI I/037/08/0, ASI I/058/10/0, ASI 2014-025-R.0, ASI 2014-025-R.1.2015, ASI 2018-24-HH.0, dedicated to the Italian participation to the Gaia Data Analysis and Processing Consortium (DPAC). This research was (partially) funded by the United Kingdom Science and Technology Facilities Council (STFC) and the United Kingdom Space Agency (UKSA) through the following grants to the University of Cambridge: ST/K000756/1, ST/N000641/1 and ST/S000089/1. This work was (partially) supported by the Spanish Ministry of Science, Innovation and University (MICIU/FEDER, UE) through grant RTI2018-095076-B-C21, and the Institute of Cosmos Sciences University of Barcelona (ICCUB, Unidad de Excelencia ’Mar\'{\i}a de Maeztu’) through grant CEX2019-000918-M. 

We used \textsc{iraf}, now not supported anymore, distributed by the National Optical Astronomy Observatories, which are operated by the Association of Universities for Research in Astronomy, Inc., under cooperative agreement with the National Science Foundation. This research has made  use of the Simbad astronomical data base \citep{simbad} and the Vizier catalogue access tool \citep{vizier}, both operated at the Centre de Donn\'ees astronomiques de Strasbourg (CDS), and of \textsc{topcat} \citep{topcat}. Figures were produced with the \textsc{r} programming language \citep{R,data.table} and \textsc{rstudio} (\url{https://www.rstudio.com/}). We also made extensive use of \textsc{python} \citep[][\url{http://www.python.org}]{van1995python} and \textsc{molecfit} \citep{molecfit, molecfit2} for the telluric features removal.

\bibliographystyle{mnras}
\bibliography{spssv2} 

\begin{thebibliography}{}
\makeatletter
\relax
\def\mn@urlcharsother{\let\do\@makeother \do\$\do\&\do\#\do\^\do\_\do\%\do\~}
\def\mn@doi{\begingroup\mn@urlcharsother \@ifnextchar [ {\mn@doi@}
  {\mn@doi@[]}}
\def\mn@doi@[#1]#2{\def\@tempa{#1}\ifx\@tempa\@empty \href
  {http://dx.doi.org/#2} {doi:#2}\else \href {http://dx.doi.org/#2} {#1}\fi
  \endgroup}
\def\mn@eprint#1#2{\mn@eprint@#1:#2::\@nil}
\def\mn@eprint@arXiv#1{\href {http://arxiv.org/abs/#1} {{\tt arXiv:#1}}}
\def\mn@eprint@dblp#1{\href {http://dblp.uni-trier.de/rec/bibtex/#1.xml}
  {dblp:#1}}
\def\mn@eprint@#1:#2:#3:#4\@nil{\def\@tempa {#1}\def\@tempb {#2}\def\@tempc
  {#3}\ifx \@tempc \@empty \let \@tempc \@tempb \let \@tempb \@tempa \fi \ifx
  \@tempb \@empty \def\@tempb {arXiv}\fi \@ifundefined
  {mn@eprint@\@tempb}{\@tempb:\@tempc}{\expandafter \expandafter \csname
  mn@eprint@\@tempb\endcsname \expandafter{\@tempc}}}

\bibitem[\protect\citeauthoryear{{Albert} et~al.,}{{Albert}
  et~al.}{2015}]{albert15}
{Albert} J.,  et~al., 2015, in American Astronomical Society Meeting Abstracts
  \#225. p. 328.07

\bibitem[\protect\citeauthoryear{{Albrecht} et~al.,}{{Albrecht}
  et~al.}{2006}]{albrecht06}
{Albrecht} A.,  et~al., 2006, arXiv e-prints, \href
  {https://ui.adsabs.harvard.edu/abs/2006astro.ph..9591A} {pp
  astro--ph/0609591}

\bibitem[\protect\citeauthoryear{{Allard}, {Homeier}  \& {Freytag}}{{Allard}
  et~al.}{2012}]{allard12}
{Allard} F.,  {Homeier} D.,   {Freytag} B.,  2012, \mn@doi [Philosophical
  Transactions of the Royal Society of London Series A]
  {10.1098/rsta.2011.0269}, \href
  {https://ui.adsabs.harvard.edu/abs/2012RSPTA.370.2765A} {370, 2765}

\bibitem[\protect\citeauthoryear{{Altavilla} et~al.,}{{Altavilla}
  et~al.}{2015}]{altavilla15}
{Altavilla} G.,  et~al., 2015, \mn@doi [Astronomische Nachrichten]
  {10.1002/asna.201512176}, \href
  {https://ui.adsabs.harvard.edu/abs/2015AN....336..515A} {336, 515}

\bibitem[\protect\citeauthoryear{{Altavilla} et~al.,}{{Altavilla}
  et~al.}{2021}]{altavilla20}
{Altavilla} G.,  et~al., 2021, \mn@doi [\mnras] {10.1093/mnras/staa3655}, \href
  {https://ui.adsabs.harvard.edu/abs/2021MNRAS.501.2848A} {501, 2848}

\bibitem[\protect\citeauthoryear{{Andrae} et~al.,}{{Andrae}
  et~al.}{2018}]{andrae18}
{Andrae} R.,  et~al., 2018, \mn@doi [\aap] {10.1051/0004-6361/201732516}, \href
  {https://ui.adsabs.harvard.edu/abs/2018A&A...616A...8A} {616, A8}

\bibitem[\protect\citeauthoryear{{Artigau}}{{Artigau}}{2018}]{artigau18}
{Artigau} {\'E}.,  2018, in {Deeg} H.~J.,  {Belmonte} J.~A.,  eds, Handbook of
  Exoplanets. p.~94, \mn@doi{10.1007/978-3-319-55333-7_94}

\bibitem[\protect\citeauthoryear{{Bailer-Jones} et~al.,}{{Bailer-Jones}
  et~al.}{2013}]{apsis}
{Bailer-Jones} C.~A.~L.,  et~al., 2013, \mn@doi [\aap]
  {10.1051/0004-6361/201322344}, \href
  {https://ui.adsabs.harvard.edu/abs/2013A&A...559A..74B} {559, A74}

\bibitem[\protect\citeauthoryear{{Baldwin} \& {Stone}}{{Baldwin} \&
  {Stone}}{1984}]{baldwin84}
{Baldwin} J.~A.,  {Stone} R.~P.~S.,  1984, \mn@doi [\mnras]
  {10.1093/mnras/206.2.241}, \href
  {https://ui.adsabs.harvard.edu/abs/1984MNRAS.206..241B} {206, 241}

\bibitem[\protect\citeauthoryear{{Bessell}}{{Bessell}}{2005}]{bessell05}
{Bessell} M.~S.,  2005, \mn@doi [\araa]
  {10.1146/annurev.astro.41.082801.100251}, \href
  {https://ui.adsabs.harvard.edu/abs/2005ARA&A..43..293B} {43, 293}

\bibitem[\protect\citeauthoryear{{Bessell} \& {Murphy}}{{Bessell} \&
  {Murphy}}{2012}]{bessell12}
{Bessell} M.,  {Murphy} S.,  2012, \mn@doi [\pasp] {10.1086/664083}, \href
  {https://ui.adsabs.harvard.edu/abs/2012PASP..124..140B} {124, 140}

\bibitem[\protect\citeauthoryear{{Blackwell}, {Leggett}, {Petford}, {Mountain}
  \& {Selby}}{{Blackwell} et~al.}{1983}]{blackwell83}
{Blackwell} D.~E.,  {Leggett} S.~K.,  {Petford} A.~D.,  {Mountain} C.~M.,
  {Selby} M.~J.,  1983, \mn@doi [\mnras] {10.1093/mnras/205.3.897}, \href
  {https://ui.adsabs.harvard.edu/abs/1983MNRAS.205..897B} {205, 897}

\bibitem[\protect\citeauthoryear{{Bohlin}}{{Bohlin}}{2014}]{bohlin14}
{Bohlin} R.~C.,  2014, \mn@doi [\aj] {10.1088/0004-6256/147/6/127}, \href
  {https://ui.adsabs.harvard.edu/abs/2014AJ....147..127B} {147, 127}

\bibitem[\protect\citeauthoryear{{Bohlin} \& {Landolt}}{{Bohlin} \&
  {Landolt}}{2015}]{bohlin15}
{Bohlin} R.~C.,  {Landolt} A.~U.,  2015, \mn@doi [\aj]
  {10.1088/0004-6256/149/4/122}, \href
  {https://ui.adsabs.harvard.edu/abs/2015AJ....149..122B} {149, 122}

\bibitem[\protect\citeauthoryear{{Bohlin}, {Colina}  \& {Finley}}{{Bohlin}
  et~al.}{1995}]{bohlin95}
{Bohlin} R.~C.,  {Colina} L.,   {Finley} D.~S.,  1995, \mn@doi [\aj]
  {10.1086/117606}, \href
  {https://ui.adsabs.harvard.edu/abs/1995AJ....110.1316B} {110, 1316}

\bibitem[\protect\citeauthoryear{{Bohlin}, {Dickinson}  \& {Calzetti}}{{Bohlin}
  et~al.}{2001}]{bohlin01}
{Bohlin} R.~C.,  {Dickinson} M.~E.,   {Calzetti} D.,  2001, \mn@doi [\aj]
  {10.1086/323137}, \href
  {https://ui.adsabs.harvard.edu/abs/2001AJ....122.2118B} {122, 2118}

\bibitem[\protect\citeauthoryear{{Bohlin}, {Deustua}  \& {de Rosa}}{{Bohlin}
  et~al.}{2019}]{bohlin19}
{Bohlin} R.~C.,  {Deustua} S.~E.,   {de Rosa} G.,  2019, \mn@doi [\aj]
  {10.3847/1538-3881/ab480c}, \href
  {https://ui.adsabs.harvard.edu/abs/2019AJ....158..211B} {158, 211}

\bibitem[\protect\citeauthoryear{{Bohlin}, {Hubeny}  \& {Rauch}}{{Bohlin}
  et~al.}{2020}]{bohlin20}
{Bohlin} R.~C.,  {Hubeny} I.,   {Rauch} T.,  2020, \mn@doi [\aj]
  {10.3847/1538-3881/ab94b4}, \href
  {https://ui.adsabs.harvard.edu/abs/2020AJ....160...21B} {160, 21}

\bibitem[\protect\citeauthoryear{{Casagrande} \& {VandenBerg}}{{Casagrande} \&
  {VandenBerg}}{2018}]{casagrande18}
{Casagrande} L.,  {VandenBerg} D.~A.,  2018, \mn@doi [\mnras]
  {10.1093/mnrasl/sly104}, \href
  {https://ui.adsabs.harvard.edu/abs/2018MNRAS.479L.102C} {479, L102}

\bibitem[\protect\citeauthoryear{{Clem} \& {Landolt}}{{Clem} \&
  {Landolt}}{2013}]{clem13}
{Clem} J.~L.,  {Landolt} A.~U.,  2013, \mn@doi [\aj]
  {10.1088/0004-6256/146/4/88}, \href
  {https://ui.adsabs.harvard.edu/abs/2013AJ....146...88C} {146, 88}

\bibitem[\protect\citeauthoryear{{Coelho}}{{Coelho}}{2014}]{coelho14}
{Coelho} P.~R.~T.,  2014, \mn@doi [\mnras] {10.1093/mnras/stu365}, \href
  {https://ui.adsabs.harvard.edu/abs/2014MNRAS.440.1027C} {440, 1027}

\bibitem[\protect\citeauthoryear{{De Marco} \& {Izzard}}{{De Marco} \&
  {Izzard}}{2017}]{demarco17}
{De Marco} O.,  {Izzard} R.~G.,  2017, \mn@doi [\pasa] {10.1017/pasa.2016.52},
  \href {https://ui.adsabs.harvard.edu/abs/2017PASA...34....1D} {34, e001}

\bibitem[\protect\citeauthoryear{{Deustua}, {Kent}  \& {Smith}}{{Deustua}
  et~al.}{2013}]{deustua13}
{Deustua} S.,  {Kent} S.,   {Smith} J.~A.,  2013, in {Oswalt} T.~D.,  {Bond}
  H.~E.,  eds, Planets, Stars and Stellar Systems. Volume 2: Astronomical
  Techniques, Software and Data. Springer, Dordrecht, p.~375,
  \mn@doi{10.1007/978-94-007-5618-2_8}

\bibitem[\protect\citeauthoryear{Dowle \& Srinivasan}{Dowle \&
  Srinivasan}{2019}]{data.table}
Dowle M.,  Srinivasan A.,  2019, data.table: Extension of `data.frame`.
\url {https://CRAN.R-project.org/package=data.table}

\bibitem[\protect\citeauthoryear{{Evans} et~al.,}{{Evans}
  et~al.}{2018}]{evans18}
{Evans} D.~W.,  et~al., 2018, \mn@doi [\aap] {10.1051/0004-6361/201832756},
  \href {https://ui.adsabs.harvard.edu/abs/2018A&A...616A...4E} {616, A4}

\bibitem[\protect\citeauthoryear{{Gaia Collaboration} et~al.,}{{Gaia
  Collaboration} et~al.}{2016a}]{gaia}
{Gaia Collaboration} et~al., 2016a, \mn@doi [\aap]
  {10.1051/0004-6361/201629272}, \href
  {https://ui.adsabs.harvard.edu/abs/2016A&A...595A...1G} {595, A1}

\bibitem[\protect\citeauthoryear{{Gaia Collaboration} et~al.,}{{Gaia
  Collaboration} et~al.}{2016b}]{gdr1}
{Gaia Collaboration} et~al., 2016b, \mn@doi [\aap]
  {10.1051/0004-6361/201629512}, \href
  {https://ui.adsabs.harvard.edu/abs/2016A&A...595A...2G} {595, A2}

\bibitem[\protect\citeauthoryear{{Gaia Collaboration} et~al.,}{{Gaia
  Collaboration} et~al.}{2018}]{gdr2}
{Gaia Collaboration} et~al., 2018, \mn@doi [\aap]
  {10.1051/0004-6361/201833051}, \href
  {https://ui.adsabs.harvard.edu/abs/2018A&A...616A...1G} {616, A1}

\bibitem[\protect\citeauthoryear{{Gaia Collaboration} et~al.,}{{Gaia
  Collaboration} et~al.}{2019}]{eyer19}
{Gaia Collaboration} et~al., 2019, \mn@doi [\aap]
  {10.1051/0004-6361/201833304}, \href
  {https://ui.adsabs.harvard.edu/abs/2019A&A...623A.110G} {623, A110}

\bibitem[\protect\citeauthoryear{{Gaia Collaboration}, {Brown}, {Vallenari},
  {Prusti}, {de Bruijne}, {Babusiaux}  \& {Biermann}}{{Gaia Collaboration}
  et~al.}{2020}]{egdr3}
{Gaia Collaboration} {Brown} A.~G.~A.,  {Vallenari} A.,  {Prusti} T.,  {de
  Bruijne} J.~H.~J.,  {Babusiaux} C.,   {Biermann} M.,  2020, arXiv e-prints,
  \href {https://ui.adsabs.harvard.edu/abs/2020arXiv201201533G} {p.
  arXiv:2012.01533}

\bibitem[\protect\citeauthoryear{{Gilmore}}{{Gilmore}}{2018}]{gilmore18}
{Gilmore} G.,  2018, \mn@doi [Contemporary Physics]
  {10.1080/00107514.2018.1439700}, \href
  {https://ui.adsabs.harvard.edu/abs/2018ConPh..59..155G} {59, 155}

\bibitem[\protect\citeauthoryear{{Hamuy}, {Suntzeff}, {Heathcote}, {Walker},
  {Gigoux}  \& {Phillips}}{{Hamuy} et~al.}{1994}]{hamuy94}
{Hamuy} M.,  {Suntzeff} N.~B.,  {Heathcote} S.~R.,  {Walker} A.~R.,  {Gigoux}
  P.,   {Phillips} M.~M.,  1994, \mn@doi [\pasp] {10.1086/133417}, \href
  {https://ui.adsabs.harvard.edu/abs/1994PASP..106..566H} {106, 566}

\bibitem[\protect\citeauthoryear{Harrison}{Harrison}{2011}]{harrison11}
Harrison D.~L.,  2011, \mn@doi [EXP ASTRON] {10.1007/s10686-011-9240-7}, 31,
  157

\bibitem[\protect\citeauthoryear{{Hermes}, {G{\"a}nsicke}, {Gentile Fusillo},
  {Raddi}, {Hollands}, {Dennihy}, {Fuchs}  \& {Redfield}}{{Hermes}
  et~al.}{2017}]{hermes17}
{Hermes} J.~J.,  {G{\"a}nsicke} B.~T.,  {Gentile Fusillo} N.~P.,  {Raddi} R.,
  {Hollands} M.~A.,  {Dennihy} E.,  {Fuchs} J.~T.,   {Redfield} S.,  2017,
  \mn@doi [\mnras] {10.1093/mnras/stx567}, \href
  {https://ui.adsabs.harvard.edu/abs/2017MNRAS.468.1946H} {468, 1946}

\bibitem[\protect\citeauthoryear{{Holl} et~al.,}{{Holl} et~al.}{2018}]{holl18}
{Holl} B.,  et~al., 2018, \mn@doi [\aap] {10.1051/0004-6361/201832892}, \href
  {https://ui.adsabs.harvard.edu/abs/2018A&A...618A..30H} {618, A30}

\bibitem[\protect\citeauthoryear{{Hosey}, {Henry}, {Jao}, {Dieterich},
  {Winters}, {Lurie}, {Riedel}  \& {Subasavage}}{{Hosey}
  et~al.}{2015}]{hosey15}
{Hosey} A.~D.,  {Henry} T.~J.,  {Jao} W.-C.,  {Dieterich} S.~B.,  {Winters}
  J.~G.,  {Lurie} J.~C.,  {Riedel} A.~R.,   {Subasavage} J.~P.,  2015, \mn@doi
  [\aj] {10.1088/0004-6256/150/1/6}, \href
  {https://ui.adsabs.harvard.edu/abs/2015AJ....150....6H} {150, 6}

\bibitem[\protect\citeauthoryear{{Husser}, {Wende-von Berg}, {Dreizler},
  {Homeier}, {Reiners}, {Barman}  \& {Hauschildt}}{{Husser}
  et~al.}{2013}]{husser13}
{Husser} T.~O.,  {Wende-von Berg} S.,  {Dreizler} S.,  {Homeier} D.,  {Reiners}
  A.,  {Barman} T.,   {Hauschildt} P.~H.,  2013, \mn@doi [\aap]
  {10.1051/0004-6361/201219058}, \href
  {https://ui.adsabs.harvard.edu/abs/2013A&A...553A...6H} {553, A6}

\bibitem[\protect\citeauthoryear{{Kaiser} et~al.,}{{Kaiser}
  et~al.}{2017}]{kaiser17}
{Kaiser} M.~E.,  et~al., 2017, in Society of Photo-Optical Instrumentation
  Engineers (SPIE) Conference Series. p. 1039815, \mn@doi{10.1117/12.2274637}

\bibitem[\protect\citeauthoryear{{Katz} et~al.,}{{Katz} et~al.}{2019}]{katz19}
{Katz} D.,  et~al., 2019, \mn@doi [\aap] {10.1051/0004-6361/201833273}, \href
  {https://ui.adsabs.harvard.edu/abs/2019A&A...622A.205K} {622, A205}

\bibitem[\protect\citeauthoryear{{Kausch} et~al.,}{{Kausch}
  et~al.}{2015}]{molecfit2}
{Kausch} W.,  et~al., 2015, \mn@doi [\aap] {10.1051/0004-6361/201423909}, \href
  {https://ui.adsabs.harvard.edu/abs/2015A&A...576A..78K} {576, A78}

\bibitem[\protect\citeauthoryear{{Khan} \& {Worthey}}{{Khan} \&
  {Worthey}}{2018}]{khan18}
{Khan} I.,  {Worthey} G.,  2018, \mn@doi [\aap] {10.1051/0004-6361/201732545},
  \href {https://ui.adsabs.harvard.edu/abs/2018A&A...615A.115K} {615, A115}

\bibitem[\protect\citeauthoryear{{Koen}, {Kilkenny}, {van Wyk}  \&
  {Marang}}{{Koen} et~al.}{2010}]{koen10}
{Koen} C.,  {Kilkenny} D.,  {van Wyk} F.,   {Marang} F.,  2010, \mn@doi
  [\mnras] {10.1111/j.1365-2966.2009.16182.x}, \href
  {https://ui.adsabs.harvard.edu/abs/2010MNRAS.403.1949K} {403, 1949}

\bibitem[\protect\citeauthoryear{{Koester}}{{Koester}}{2010}]{koester10}
{Koester} D.,  2010, \memsai, \href
  {https://ui.adsabs.harvard.edu/abs/2010MmSAI..81..921K} {81, 921}

\bibitem[\protect\citeauthoryear{{Koleva} \& {Vazdekis}}{{Koleva} \&
  {Vazdekis}}{2012}]{ngsl}
{Koleva} M.,  {Vazdekis} A.,  2012, \mn@doi [\aap]
  {10.1051/0004-6361/201118065}, \href
  {https://ui.adsabs.harvard.edu/abs/2012A&A...538A.143K} {538, A143}

\bibitem[\protect\citeauthoryear{{Landolt}}{{Landolt}}{1983}]{landolt83}
{Landolt} A.~U.,  1983, \mn@doi [\aj] {10.1086/113329}, \href
  {https://ui.adsabs.harvard.edu/abs/1983AJ.....88..439L} {88, 439}

\bibitem[\protect\citeauthoryear{{Landolt}}{{Landolt}}{1992a}]{landolt92}
{Landolt} A.~U.,  1992a, \mn@doi [\aj] {10.1086/116242}, \href
  {https://ui.adsabs.harvard.edu/abs/1992AJ....104..340L} {104, 340}

\bibitem[\protect\citeauthoryear{{Landolt}}{{Landolt}}{1992b}]{landolt92b}
{Landolt} A.~U.,  1992b, \mn@doi [\aj] {10.1086/116243}, \href
  {https://ui.adsabs.harvard.edu/abs/1992AJ....104..372L} {104, 372}

\bibitem[\protect\citeauthoryear{{Landolt}}{{Landolt}}{2009}]{landolt09}
{Landolt} A.~U.,  2009, \mn@doi [\aj] {10.1088/0004-6256/137/5/4186}, \href
  {https://ui.adsabs.harvard.edu/abs/2009AJ....137.4186L} {137, 4186}

\bibitem[\protect\citeauthoryear{{Landolt} \& {Uomoto}}{{Landolt} \&
  {Uomoto}}{2007}]{landolt07}
{Landolt} A.~U.,  {Uomoto} A.~K.,  2007, \mn@doi [\aj] {10.1086/510485}, \href
  {https://ui.adsabs.harvard.edu/abs/2007AJ....133..768L} {133, 768}

\bibitem[\protect\citeauthoryear{{Levenhagen}, {Diaz}, {Coelho}  \&
  {Hubeny}}{{Levenhagen} et~al.}{2017}]{levenhagen17}
{Levenhagen} R.~S.,  {Diaz} M.~P.,  {Coelho} P. R.~T.,   {Hubeny} I.,  2017,
  \mn@doi [\apjs] {10.3847/1538-4365/aa7681}, \href
  {https://ui.adsabs.harvard.edu/abs/2017ApJS..231....1L} {231, 1}

\bibitem[\protect\citeauthoryear{{Lindegren} et~al.,}{{Lindegren}
  et~al.}{2020}]{lindegren20}
{Lindegren} L.,  et~al., 2020, arXiv e-prints, \href
  {https://ui.adsabs.harvard.edu/abs/2020arXiv201203380L} {p. arXiv:2012.03380}

\bibitem[\protect\citeauthoryear{{MacGregor}, {Weinberger}, {Wilner},
  {Kowalski}  \& {Cranmer}}{{MacGregor} et~al.}{2018}]{macgregor18}
{MacGregor} M.~A.,  {Weinberger} A.~J.,  {Wilner} D.~J.,  {Kowalski} A.~F.,
  {Cranmer} S.~R.,  2018, \mn@doi [\apjl] {10.3847/2041-8213/aaad6b}, \href
  {https://ui.adsabs.harvard.edu/abs/2018ApJ...855L...2M} {855, L2}

\bibitem[\protect\citeauthoryear{{Ma{\'\i}z Apell{\'a}niz} \&
  {Weiler}}{{Ma{\'\i}z Apell{\'a}niz} \& {Weiler}}{2018}]{maiz18}
{Ma{\'\i}z Apell{\'a}niz} J.,  {Weiler} M.,  2018, \mn@doi [\aap]
  {10.1051/0004-6361/201834051}, \href
  {https://ui.adsabs.harvard.edu/abs/2018A&A...619A.180M} {619, A180}

\bibitem[\protect\citeauthoryear{{Malumuth}, {Hill}, {Cheng}, {Cottingham},
  {Wen}, {Johnson}  \& {Hill}}{{Malumuth} et~al.}{2003}]{malumuth03}
{Malumuth} E.~M.,  {Hill} R.~J.,  {Cheng} E.~S.,  {Cottingham} D.~A.,  {Wen}
  Y.,  {Johnson} S.~D.,   {Hill} R.~S.,  2003, in {Blades} J.~C.,  {Siegmund}
  O. H.~W.,  eds,  Society of Photo-Optical Instrumentation Engineers (SPIE)
  Conference Series Vol. 4854, Future EUV/UV and Visible Space Astrophysics
  Missions and Instrumentation.. pp 567--576, \mn@doi{10.1117/12.459789}

\bibitem[\protect\citeauthoryear{{Marinoni} et~al.,}{{Marinoni}
  et~al.}{2016}]{marinoni16}
{Marinoni} S.,  et~al., 2016, \mn@doi [\mnras] {10.1093/mnras/stw1886}, \href
  {https://ui.adsabs.harvard.edu/abs/2016MNRAS.462.3616M} {462, 3616}

\bibitem[\protect\citeauthoryear{{Martin}, {Zapatero Osorio}  \&
  {Rebolo}}{{Martin} et~al.}{1996}]{martin96}
{Martin} E.~L.,  {Zapatero Osorio} M.~R.,   {Rebolo} R.,  1996, in
  {Pallavicini} R.,  {Dupree} A.~K.,  eds,  Astronomical Society of the Pacific
  Conference Series Vol. 109, Cool Stars, Stellar Systems, and the Sun. p.~615

\bibitem[\protect\citeauthoryear{{Megessier}}{{Megessier}}{1995}]{megessier95}
{Megessier} C.,  1995, \aap, \href
  {https://ui.adsabs.harvard.edu/abs/1995A&A...296..771M} {296, 771}

\bibitem[\protect\citeauthoryear{{Mikhailenko}, {Babikov}  \&
  {Golovko}}{{Mikhailenko} et~al.}{2005}]{spectra}
{Mikhailenko} S.~N.,  {Babikov} Y.~L.,   {Golovko} V.~F.,  2005, Atmospheric
  and oceanic optics, V18, N.09, 685

\bibitem[\protect\citeauthoryear{{Moehler} et~al.,}{{Moehler}
  et~al.}{2014}]{moehler14}
{Moehler} S.,  et~al., 2014, \mn@doi [\aap] {10.1051/0004-6361/201423790},
  \href {https://ui.adsabs.harvard.edu/abs/2014A&A...568A...9M} {568, A9}

\bibitem[\protect\citeauthoryear{{Mowlavi} et~al.,}{{Mowlavi}
  et~al.}{2018}]{mowlavi18}
{Mowlavi} N.,  et~al., 2018, \mn@doi [\aap] {10.1051/0004-6361/201833366},
  \href {https://ui.adsabs.harvard.edu/abs/2018A&A...618A..58M} {618, A58}

\bibitem[\protect\citeauthoryear{{Ochsenbein}, {Bauer}  \&
  {Marcout}}{{Ochsenbein} et~al.}{2000}]{vizier}
{Ochsenbein} F.,  {Bauer} P.,   {Marcout} J.,  2000, \mn@doi [\aaps]
  {10.1051/aas:2000169}, \href
  {https://ui.adsabs.harvard.edu/abs/2000A&AS..143...23O} {143, 23}

\bibitem[\protect\citeauthoryear{{Palacios}, {Gebran}, {Josselin}, {Martins},
  {Plez}, {Belmas}  \& {L{\`e}bre}}{{Palacios} et~al.}{2010}]{palacios10}
{Palacios} A.,  {Gebran} M.,  {Josselin} E.,  {Martins} F.,  {Plez} B.,
  {Belmas} M.,   {L{\`e}bre} A.,  2010, \mn@doi [\aap]
  {10.1051/0004-6361/200913932}, \href
  {https://ui.adsabs.harvard.edu/abs/2010A&A...516A..13P} {516, A13}

\bibitem[\protect\citeauthoryear{{Pancino}}{{Pancino}}{2020}]{pancino20}
{Pancino} E.,  2020, \mn@doi [Advances in Space Research]
  {10.1016/j.asr.2019.11.007}, \href
  {https://ui.adsabs.harvard.edu/abs/2020AdSpR..65....1P} {65, 1}

\bibitem[\protect\citeauthoryear{{Pancino} et~al.,}{{Pancino}
  et~al.}{2012}]{pancino12}
{Pancino} E.,  et~al., 2012, \mn@doi [\mnras]
  {10.1111/j.1365-2966.2012.21766.x}, \href
  {https://ui.adsabs.harvard.edu/abs/2012MNRAS.426.1767P} {426, 1767}

\bibitem[\protect\citeauthoryear{{Perryman} et~al.,}{{Perryman}
  et~al.}{2001}]{perryman01}
{Perryman} M.~A.~C.,  et~al., 2001, \mn@doi [\aap]
  {10.1051/0004-6361:20010085}, \href
  {https://ui.adsabs.harvard.edu/abs/2001A&A...369..339P} {369, 339}

\bibitem[\protect\citeauthoryear{{Price}, {Paxson}, {Engelke}  \&
  {Murdock}}{{Price} et~al.}{2004}]{price04}
{Price} S.~D.,  {Paxson} C.,  {Engelke} C.,   {Murdock} T.~L.,  2004, \mn@doi
  [\aj] {10.1086/422024}, \href
  {https://ui.adsabs.harvard.edu/abs/2004AJ....128..889P} {128, 889}

\bibitem[\protect\citeauthoryear{{R Core Team}}{{R Core Team}}{2018}]{R}
{R Core Team} 2018, R: A Language and Environment for Statistical Computing.
R Foundation for Statistical Computing, Vienna, Austria, \url
  {https://www.R-project.org/}

\bibitem[\protect\citeauthoryear{{Rauch}, {Rudkowski}, {Kampka}, {Werner},
  {Kruk}  \& {Moehler}}{{Rauch} et~al.}{2014}]{rauch14}
{Rauch} T.,  {Rudkowski} A.,  {Kampka} D.,  {Werner} K.,  {Kruk} J.~W.,
  {Moehler} S.,  2014, \mn@doi [\aap] {10.1051/0004-6361/201423711}, \href
  {https://ui.adsabs.harvard.edu/abs/2014A&A...566A...3R} {566, A3}

\bibitem[\protect\citeauthoryear{{Ribas}, {Gregg}, {Boyajian}  \&
  {Bolmont}}{{Ribas} et~al.}{2017}]{ribas17}
{Ribas} I.,  {Gregg} M.~D.,  {Boyajian} T.~S.,   {Bolmont} E.,  2017, \mn@doi
  [\aap] {10.1051/0004-6361/201730582}, \href
  {https://ui.adsabs.harvard.edu/abs/2017A&A...603A..58R} {603, A58}

\bibitem[\protect\citeauthoryear{{Riello} et~al.,}{{Riello}
  et~al.}{2020}]{riello20}
{Riello} M.,  et~al., 2020, arXiv e-prints, \href
  {https://ui.adsabs.harvard.edu/abs/2020arXiv201201916R} {p. arXiv:2012.01916}

\bibitem[\protect\citeauthoryear{{Rothman} et~al.,}{{Rothman}
  et~al.}{2009}]{hitran}
{Rothman} L.~S.,  et~al., 2009, \mn@doi [\jqsrt] {10.1016/j.jqsrt.2009.02.013},
  \href {https://ui.adsabs.harvard.edu/abs/2009JQSRT.110..533R} {110, 533}

\bibitem[\protect\citeauthoryear{{S{\'a}nchez-Bl{\'a}zquez}
  et~al.,}{{S{\'a}nchez-Bl{\'a}zquez} et~al.}{2006}]{sanchez06}
{S{\'a}nchez-Bl{\'a}zquez} P.,  et~al., 2006, \mn@doi [\mnras]
  {10.1111/j.1365-2966.2006.10699.x}, \href
  {https://ui.adsabs.harvard.edu/abs/2006MNRAS.371..703S} {371, 703}

\bibitem[\protect\citeauthoryear{{Schwartz} \& {Melnik}}{{Schwartz} \&
  {Melnik}}{1993}]{eso93}
{Schwartz} H.~E.,  {Melnik} J.,  1993, The ESO users manual.
European Southern Observatory, Garching, Germany

\bibitem[\protect\citeauthoryear{{Selby}, {Mountain}, {Blackwell}, {Petford}
  \& {Leggett}}{{Selby} et~al.}{1983}]{selby83}
{Selby} M.~J.,  {Mountain} C.~M.,  {Blackwell} D.~E.,  {Petford} A.~D.,
  {Leggett} S.~K.,  1983, \mn@doi [\mnras] {10.1093/mnras/203.3.795}, \href
  {https://ui.adsabs.harvard.edu/abs/1983MNRAS.203..795S} {203, 795}

\bibitem[\protect\citeauthoryear{{Smette} et~al.,}{{Smette}
  et~al.}{2015}]{molecfit}
{Smette} A.,  et~al., 2015, \mn@doi [\aap] {10.1051/0004-6361/201423932}, \href
  {https://ui.adsabs.harvard.edu/abs/2015A&A...576A..77S} {576, A77}

\bibitem[\protect\citeauthoryear{{Smith}, {Woodward}, {Jenkins}, {Brown}  \&
  {Lykke}}{{Smith} et~al.}{2009}]{smith09}
{Smith} A.~W.,  {Woodward} J.~T.,  {Jenkins} C.~A.,  {Brown} S.~W.,   {Lykke}
  K.~R.,  2009, \mn@doi [Metrologia] {10.1088/0026-1394/46/4/S16}, \href
  {https://ui.adsabs.harvard.edu/abs/2009Metro..46S.219S} {46, S219}

\bibitem[\protect\citeauthoryear{{Sordo} et~al.,}{{Sordo}
  et~al.}{2011}]{sordo11}
{Sordo} R.,  et~al., 2011, in Journal of Physics Conference Series. p. 012006,
  \mn@doi{10.1088/1742-6596/328/1/012006}

\bibitem[\protect\citeauthoryear{{Soubiran} et~al.,}{{Soubiran}
  et~al.}{2018}]{soubiran18}
{Soubiran} C.,  et~al., 2018, \mn@doi [\aap] {10.1051/0004-6361/201832795},
  \href {https://ui.adsabs.harvard.edu/abs/2018A&A...616A...7S} {616, A7}

\bibitem[\protect\citeauthoryear{{Stetson}}{{Stetson}}{2000}]{stetson00}
{Stetson} P.~B.,  2000, \mn@doi [\pasp] {10.1086/316595}, \href
  {https://ui.adsabs.harvard.edu/abs/2000PASP..112..925S} {112, 925}

\bibitem[\protect\citeauthoryear{{Stetson}, {Pancino}, {Zocchi}, {Sanna}  \&
  {Monelli}}{{Stetson} et~al.}{2019}]{stetson19}
{Stetson} P.~B.,  {Pancino} E.,  {Zocchi} A.,  {Sanna} N.,   {Monelli} M.,
  2019, \mn@doi [\mnras] {10.1093/mnras/stz585}, \href
  {https://ui.adsabs.harvard.edu/abs/2019MNRAS.485.3042S} {485, 3042}

\bibitem[\protect\citeauthoryear{{Stone} \& {Baldwin}}{{Stone} \&
  {Baldwin}}{1983}]{stone83}
{Stone} R.~P.~S.,  {Baldwin} J.~A.,  1983, \mn@doi [\mnras]
  {10.1093/mnras/204.2.347}, \href
  {https://ui.adsabs.harvard.edu/abs/1983MNRAS.204..347S} {204, 347}

\bibitem[\protect\citeauthoryear{{Strai{\v{z}}ys}}{{Strai{\v{z}}ys}}{1992}]{straizys92}
{Strai{\v{z}}ys} V.,  1992, {Multicolor stellar photometry}.
Pachart Pub. House, Tucson

\bibitem[\protect\citeauthoryear{{Stritzinger}, {Suntzeff}, {Hamuy}, {Challis},
  {Demarco}, {Germany}  \& {Soderberg}}{{Stritzinger}
  et~al.}{2005}]{stritzinger05}
{Stritzinger} M.,  {Suntzeff} N.~B.,  {Hamuy} M.,  {Challis} P.,  {Demarco} R.,
   {Germany} L.,   {Soderberg} A.~M.,  2005, \mn@doi [\pasp] {10.1086/431468},
  \href {https://ui.adsabs.harvard.edu/abs/2005PASP..117..810S} {117, 810}

\bibitem[\protect\citeauthoryear{{Taylor}}{{Taylor}}{2005}]{topcat}
{Taylor} M.~B.,  2005, in {Shopbell} P.,  {Britton} M.,   {Ebert} R.,  eds,
  Astronomical Society of the Pacific Conference Series Vol. 347, Astronomical
  Data Analysis Software and Systems XIV. p.~29

\bibitem[\protect\citeauthoryear{{Thuillier}, {Hers{\'e}}, {Labs}, {Foujols},
  {Peetermans}, {Gillotay}, {Simon}  \& {Mandel}}{{Thuillier}
  et~al.}{2003}]{sun}
{Thuillier} G.,  {Hers{\'e}} M.,  {Labs} D.,  {Foujols} T.,  {Peetermans} W.,
  {Gillotay} D.,  {Simon} P.~C.,   {Mandel} H.,  2003, \mn@doi [\solphys]
  {10.1023/A:1024048429145}, \href
  {https://ui.adsabs.harvard.edu/abs/2003SoPh..214....1T} {214, 1}

\bibitem[\protect\citeauthoryear{{Tody}}{{Tody}}{1986}]{irafold}
{Tody} D.,  1986, in {Crawford} D.~L.,  ed.,  Society of Photo-Optical
  Instrumentation Engineers (SPIE) Conference Series Vol. 627, Instrumentation
  in astronomy VI. p.~733, \mn@doi{10.1117/12.968154}

\bibitem[\protect\citeauthoryear{{Tody}}{{Tody}}{1993}]{iraf}
{Tody} D.,  1993, in {Hanisch} R.~J.,  {Brissenden} R.~J.~V.,   {Barnes} J.,
  eds,  Astronomical Society of the Pacific Conference Series Vol. 52,
  Astronomical Data Analysis Software and Systems II. p.~173

\bibitem[\protect\citeauthoryear{{Valdes}}{{Valdes}}{1992}]{valdes92}
{Valdes} F.,  1992, in {Worrall} D.~M.,  {Biemesderfer} C.,   {Barnes} J.,
  eds,  Astronomical Society of the Pacific Conference Series Vol. 25,
  Astronomical Data Analysis Software and Systems I. p.~417

\bibitem[\protect\citeauthoryear{Van~Rossum \& Drake~Jr}{Van~Rossum \&
  Drake~Jr}{1995}]{van1995python}
Van~Rossum G.,  Drake~Jr F.~L.,  1995, Python tutorial.
Centrum voor Wiskunde en Informatica Amsterdam, The Netherlands

\bibitem[\protect\citeauthoryear{{Weiler}}{{Weiler}}{2018}]{weiler18}
{Weiler} M.,  2018, \mn@doi [\aap] {10.1051/0004-6361/201833462}, \href
  {https://ui.adsabs.harvard.edu/abs/2018A&A...617A.138W} {617, A138}

\bibitem[\protect\citeauthoryear{{Wenger} et~al.,}{{Wenger}
  et~al.}{2000}]{simbad}
{Wenger} M.,  et~al., 2000, \mn@doi [\aaps] {10.1051/aas:2000332}, \href
  {https://ui.adsabs.harvard.edu/abs/2000A&AS..143....9W} {143, 9}

\makeatother
\end{thebibliography}


\bsp	
\label{lastpage}
\end{document}